\documentclass[useAMS,usenatbib]{mn2e}
\usepackage{graphicx}
\title[NGC~1360]{SALT HRS discovery of a long period double-degenerate binary in the planetary nebula NGC~1360\thanks{Based on observations made with the Southern African Large Telescope (SALT) under programmes 2016-2-SCI-034 and 2016-2-SCI-003, and archival data from the VLT at the Paranal Observatory under programme 60.A-9800(C).}}
\author[Miszalski et al.]{B. Miszalski,$^{1,2}$\thanks{E-mail: brent@saao.ac.za} R. Manick,$^{3}$ J. Miko{\l}ajewska,$^{4}$ K. I{\l}kiewicz,$^{4}$ D. Kamath$^{5,6,7,3}$\newauthor and H. Van Winckel$^{3}$\\
$^{1}$South African Astronomical Observatory, PO Box 9, Observatory, 7935, South Africa\\
$^{2}$Southern African Large Telescope Foundation, PO Box 9, Observatory, 7935, South Africa\\
$^{3}$Instituut voor Sterrenkunde, KU Leuven, Celestijnenlaan 200D bus 2401, B-3001 Leuven, Belgium\\
$^{4}$Nicolaus Copernicus Astronomical Center, Polish Academy of Sciences, Bartycka 18, PL-00716 Warsaw, Poland\\
$^{5}$Department of Physics and Astronomy, Macquarie University, Sydney NSW 2109, Australia\\
$^{6}$Astronomy, Astrophysics and Astrophotonics Research Centre, Macquarie University, Sydney NSW 2109, Australia\\
$^{7}$Australian Astronomical Observatory, PO Box 915, North Ryde, NSW 1670, Australia\\
}
\begin{document}

\date{Accepted . Received ; in original form }

\maketitle
\begin{abstract}
Whether planetary nebulae (PNe) are predominantly the product of binary stellar evolution as some population synthesis models (PSM) suggest remains an open question. Around 50 short period binary central stars ($P\sim1$ d) are known, but with only four with measured orbital periods over 10 d, our knowledge is severely incomplete. Here we report on the first discovery from a systematic SALT HRS survey for long period binary central stars. We find a 142 d orbital period from radial velocities of the central star of NGC~1360, HIP~16566. NGC~1360 appears to be the product of common-envelope (CE) evolution, with nebula features similar to post-CE PNe, albeit with an orbital period considerably longer than expected to be typical of post-CE PSM. The most striking feature is a newly-identified ring of candidate low-ionisation structures (LIS). Previous spatio-kinematic modelling of the nebula gives a nebula inclination of $30\pm10$ deg, and assuming the binary nucleus is coplanar with the nebula, multi-wavelength observations best fit a more massive, evolved WD companion. A WD companion in a 142 d orbit is not the focus of many PSM, making NGC~1360 a valuable system with which to improve future PSM work. HIP~16566 is amongst many central stars in which large radial velocity variability was found by low-resolution surveys. The discovery of its binary nature may indicate long period binaries may be more common than PSM models predict.
\end{abstract}


\section{Introduction}
\label{sec:intro}
Great strides have been made in recent years in the search for binary central stars of planetary nebulae (CSPNe). The previously uncertain role of binaries in shaping PNe (Balick \& Frank 2002) has been surplanted by an increasingly clear picture that binaries play a fundamental role, particularly because of new theoretical advances complemented by direct and growing evidence for binarity in PNe (De Marco 2009). Almost 50 close binaries are now known, with the majority exhibiting orbital periods of 1 d or less (Miszalski et al. 2011a; Jones 2015). The short orbital periods are the result of at least one common-envelope (CE) phase (Paczynski 1976; Ivanova et al. 2013) and together they represent the youngest accessible window into the aftermath of this critical and unobserved phase of binary stellar evolution. Around 1 in 5 PNe host a close binary (Bond 2000; Miszalski et al. 2009a) and the observed orbital period distribution is an important constraint for population synthesis models (PSM, e.g. De Kool \& Ritter 1993; Yungelson et al. 1993; Han et al. 1995; Han 1998; Moe \& De Marco 2006, 2012; Nie et al. 2012; Madappatt et al. 2016). 

While the observed orbital period distribution is consistent with that from more evolved white dwarf (WD) binaries with main-sequence companions (Rebassa-Mansergas et al. 2008; Miszalski et al. 2009a; Davis et al. 2010; Nebot G{\'o}mez-Mor{\'a}n et al. 2011), the statistics surrounding post-CE PNe remain heavily distorted by observational selection effects. The predominant selection effect is the bias towards short period systems imparted by the widespread use of photometric monitoring to identify periodic variability. Main sequence companions are readily detected via sinusoidal variability produced by irradiation of their primary-facing hemispheres, however the small amplitudes are difficult to observe for orbital periods $\ga$1 d (De Marco et al. 2008). In a limited number of cases space-based observations can circumvent this limitation (De Marco et al. 2015). Ellipsoidal variability or eclipses are also commonly observed. Double degenerates are more difficult to detect as they may not necessarily produce an irradiation effect or ellipsoidal variability.

Double degenerates and binaries with orbital periods $\ga$ 1 d are best found via the radial velocity (RV) monitoring technique. Historically, RV monitoring surveys have revealed high degrees of variability in CSPNe (e.g. M{\'e}ndez 1989; M{\'e}ndez et al. 1990; De Marco et al. 2004, 2007; Sorensen \& Pollacco 2004; Af{\v s}ar \& Bond 2005). Unfortunately, it has proven difficult to detect clear periodic signals that prove binarity. Only four close binaries have been discovered largely from RV variability or monitoring: NGC2346 ($P=15.99$ d, M{\'e}ndez \& Niemel{\"a} 1981), PN G135.9$+$55.9 ($P=0.16$ d, Tovmassian et al. 2004, 2010), Fleming~1 ($P=1.19$ d, Boffin et al. 2012) and NGC~5189 ($P=4.05$ d, Manick et al. 2015; Miszalski et al. 2015). These discoveries are notable for their inconspicuous WD companions.

Apart from close binaries, the real power of RV monitoring lies in the almost unexplored realm of long orbital periods of several weeks, months and years. The burning issue we want to resolve is whether there exists many PNe with similarly long periods and non-zero eccentricities as found in post-AGB binaries (Van Winckel 2003; Van Winckel et al. 2009). The same distribution is also observed in symbiotic stars that are the obvious successors of post-AGB binaries (Miko{\l}ajewska 2012). Population synthesis models do not currently predict the observed population of post-AGB binaries (e.g. Nie et al. 2012), but if a similar population exists in PNe, then the binary fraction for PNe could reach 40--50\% (De Marco et al. 2009). The answers to this riddle can only be revealed by long-term monitoring of CSPNe with high resolution {\'e}chelle spectroscopy. Currently only three binaries in this domain have been found with orbital periods of $P=1105\pm24$ d in BD$+$33$^\circ$~2642 (PN G052.7$+$50.7; Van Winckel et al. 2014), $P=2717\pm63$ d in HD~112313 (PN G339.9$+$88.4, LoTr~5; Van Winckel et al. 2014, Jones et al. 2017) and $P=3306\pm60$ d in BD$+$30$^\circ$623 (PN G165.5$-$15.2, NGC~1514; Jones et al. 2017). Their orbital eccentricities are 0.0, $0.26\pm0.02$ and $0.46\pm0.11$, respectively. PN G052.7$+$50.7 is an unusual metal-poor PN ionised by a very cool central star ($T_\mathrm{eff}=20$ kK, Napiwotzki et al. 1994), whereas LoTr~5 contains a fast-rotating, s-process-enhanced Barium star with a WD companion (Thevenin \& Jasniewicz 1997). NGC~1514 is remarkable for its mid-infrared rings (Ressler et al. 2010) and the nucleus shows a composite spectrum consisting of an A0 cool component and a hot subdwarf (Aller et al. 2015). 

Besides these three discoveries, it remains unclear whether a larger population of similar binary CSPNe exists. We can, however, infer that several CSPNe are highly likely to be members of this group. Barium giants have large radii and have WD companions with long orbital periods (Van der Swaelmen et al. 2017 and ref. therein). We therefore expect the three other Barium CSPNe, two of which have had WD companions detected in the UV, to have similar periods (WeBo~1, Bond et al. 2003, Siegel et al. 2012; A~70, Miszalski et al. 2012; Hen~2-39, Miszalski et al. 2013). Similarly, D'-type symbiotic stars also have red giants that show s-process enhancements and have PN-like nebulae (e.g. PC11, Pereira et al. 2010; AS201 and Cn1-1, Pereira et al. 2005, Miszalski et al. 2012; StHa190, Smith et al. 2001; V417 Cen, Gromadzki et al. 2011). Other long period binaries could be found amongst rapidly rotating red giant CSPNe that lack s-process enhancements (e.g. Me~1-1, Pereira et al. 2008; LoTr~1, Tyndall et al. 2013). Other promising candidates could be drawn from surveys looking for excess flux at X-ray (Montez et al. 2015) or near- and mid-infrared wavelengths (e.g. Bil{\'{\i}}kov{\'a} et al. 2012; De Marco et al. 2013; Douchin et al. 2015), but these efforts can be affected by chance superpositions or spatially-resolved companions (e.g. Ciardullo et al. 1999). 

A systematic RV monitoring survey is needed to target these promising long period binary candidates. Moreover, a systematic study of a large sample of other central stars is clearly required to quantify the occurrence rates of long period binaries. Here we present the first discovery of a long orbital period spectroscopic binary CSPN from an ongoing, systematic survey of a large number CSPNe with the 11-m Southern African Large Telescope (SALT, Buckley, Swart \& Meiring 2006; O'Donoghue et al. 2006). The queue-scheduled operation of SALT (V{\"a}is{\"a}nen et al. 2016) is ideally suited to long-term monitoring of a large number of CSPNe. Section \ref{sec:ngc1360} introduces HIP~16566 and its nebula NGC~1360 (PN G220.3$-$53.9). Section \ref{sec:lis} discusses the low-ionisation structures present in NGC~1360, including the discovery of a new, extensive ring of knots. Section \ref{sec:obs} describes the observations and the results are analysed in Sect. \ref{sec:analysis}. The post-CE nature of NGC~1360 is discussed in Sect. \ref{sec:postCE} and we conclude the paper in Section \ref{sec:conclusion}.

\section{HIP~16566 and NGC~1360}
\label{sec:ngc1360}
\begin{figure}
   \begin{center}
      \includegraphics[scale=0.35,angle=270]{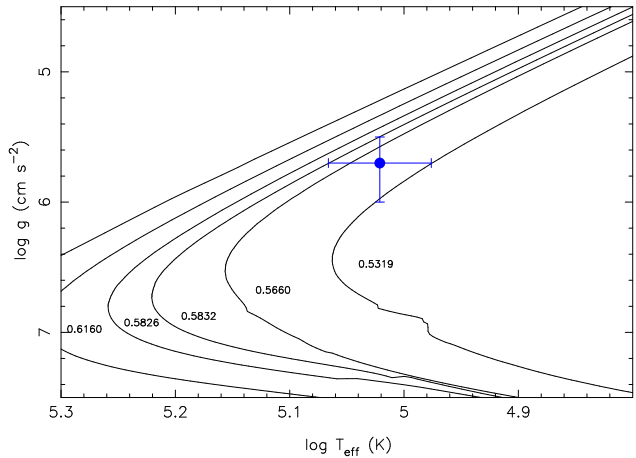}
   \end{center}
   \caption{Location of HIP~16566 in the log $T_\mathrm{eff}$-log $g$ plane with $Z=0.01$ single-star evolutionary tracks from Miller Bertolami et al. (2016).}
   \label{fig:tracks}
\end{figure}

HIP~16566 is the bright $V=11.34$ mag central star of the well-studied PN NGC~1360 (PN G220.3$-$53.9).\footnote{Also known as `The Robin's Egg' nebula.} Table \ref{tab:summary} gives a compilation of various properties of HIP~16566 from the literature. We adopt many parameters from non-local thermodynamic equilibrium (NLTE) model atmosphere fitting by Herald \& Bianchi (2011). Very similar values were also derived by Ziegler (2013). Figure \ref{fig:tracks} shows the location of HIP~16566 amongst $Z=0.01$ single-star evolutionary tracks from Miller Bertolami et al. (2016) that we have interpolated against to estimate by eye $M=0.555\pm0.030$ $M_\odot$. We calculated the gravity distance using equation 2 of M{\'e}ndez et al. (1988) with the values in Tab. \ref{tab:summary} and $F_*=12.2\times10^{16}$ erg cm$^{-2}$ s$^{-1}$ cm$^{-1}$ (M\'endez et al. 1988), resulting in $d=421_{-131}^{+123}$ pc. We adopt this distance over the more uncertain \emph{Hipparcos} distance of $\sim$350 pc (Acker et al. 1998).

\begin{table}
   \centering
   \caption{Characteristics and derived parameters of HIP~16566.}
   \label{tab:summary}
   \begin{tabular}{lrlr}
\hline
   Sp. Type & O(H) &     & (1) \\
   $V$ & 11.34 & mag & (2)\\
   $E(B-V)$ & $0.005$ & mag & (3) \\
   $T_\mathrm{eff}$ & $105\pm10$ & kK &  (3) \\
   log $g$ & $5.7^{+0.3}_{-0.2}$ & cm s$^{-2}$ & (3)  \\
   $M$ & $0.555\pm0.030$ & $M_\odot$ & this work\\
   $d$  &  $421_{-131}^{+123}$ & pc  & this work\\
   log $L/L_\odot$ & $3.5$ & & this work \\
   $\dot{M}$ & $\la$1$\times$$10^{-10}$ & $M_\odot$ yr$^{-1}$ & (3) \\
   Median $E_\mathrm{X}$ & 0.27 & keV & (4)\\
\hline
   \end{tabular}
   \begin{flushleft}
      References: (1) M{\'e}ndez (1991); (2) Ciardullo et al. (1999); (3) Herald \& Bianchi (2011); (4) Kastner et al. (2012)
   \end{flushleft}
\end{table}

There has been considerable interest in discerning the potential magnetic and binary properties of HIP~16566. The star features prominently in spectropolarimetric studies that endeavoured to detect a magnetic field. Jordan et al. (2005) first claimed to have detected a kilogauss magnetic field, but this has since been disproven with an upper limit of only a few hundred gauss now established (Leone et al. 2011, 2014; Jordan et al. 2012). M{\'e}ndez \& Niemel{\"a} (1977) claimed to have discovered binary motion with  $P=8.2$ d in HIP~16566, but Wehmeyer \& Kohoutek (1979) could not reproduce their result.  Af{\v s}ar \& Bond (2005) found significant variability on the order of 12 km s$^{-1}$ from 21 spectra of 2.2 \AA\ resolution, though no periodicity was found. Bil{\'{\i}}kov{\'a} et al. (2012) found no apparent NIR or MIR excess that could be attributed to a cool companion.

An additional window into the possible binary nature of HIP~16566 comes from X-ray observations. NGC~1360 was the first PN to be observed at X-ray wavelengths (de Korte et al. 1985; see also Guerrero et al. 2000 and ref. therein). Here we restrict our focus to high quality \emph{Chandra} observations which detected soft X-ray emission from HIP~16566 with a median energy of 0.27 keV (Kastner et al. 2012). The origin of the X-ray emission remained unconstrained in recent modelling by Montez et al. (2015), but the data can provide some constraints. Coronal emission from a main-sequence companion is unlikely as the X-rays are considerably softer than and do not overlap in energy with the three binaries studied in Montez et al. (2010) which have lower and upper quartile ranges spanning 1.01--1.13 keV (Kastner et al. 2012). Given the emission is softer than 0.3 keV, the most likely origin would be photospheric emission from HIP~16566 or a WD companion. The high-energy tail of NLTE model atmospheres may account for the soft emission (Guerrero et al. 2011). At energies higher than $\sim$0.3 keV it becomes more difficult to reproduce with NLTE models, unless the composition is H-deficient (Montez \& Kastner 2013; Montez et al. 2015). This does not apply to the H-rich HIP~16566 (Herald \& Bianchi 2011). A less likely explanation might be self-shocking winds, but the apparent weak wind of HIP~16566 in the UV (Herald \& Bianchi 2011) lead Montez et al. (2015) to consider this possibility unlikely pending more detailed investigation into self-shocking winds. On the other hand, signs of a wind are present in the optical with weak He~II emission (M\'endez et al. 1988), though this was not incorporated into the Herald \& Bianchi (2011) or Ziegler (2013) studies that used low-resolution optical spectra.

The nebula itself is notable for its large size ($9\arcmin\times7\arcmin$, Goldman et al. 2004), negligible interstellar reddening and high degree of nebular excitation (Kaler 1981). Spatio-kinematic studies of the nebula were performed by Goldman et al. (2004) and Garc{\'{\i}}a-D{\'{\i}}az et al. (2008), with both studies reaching similar conclusions. Goldman et al. (2004) found that a prolate ellipsoidal shell with an inclination of $i=30\pm10$ degrees with respect to the plane of the sky is well-constrained by the observed kinematics. The density structure was also modelled by Goldman et al. (2004) as a smooth, almost featureless thick shell with a Gaussian radial density profile. Several lower surface brightness patches are evident on the eastern side of the nebula and their presence in the MIR at 24 $\mu$m (Chu et al. 2009) suggests they can be attributed to density variations rather than local extinction effects. Based on the apparent lack of a stellar wind in UV spectra, Garc{\'{\i}}a-D{\'{\i}}az et al. (2008) proposed that the density structure of NGC~1360 was consistent with photoionised gas falling back towards the central star. Garc{\'{\i}}a-D{\'{\i}}az et al. (2008) further linked such gas motions with the observed widths of the H$\alpha$ profile being larger than the thermal width. Sch\"onberner et al. (2010) point out, however, that it is not a straightforward proposition that a smooth nebula can be created by turning off a stellar wind and that more sophisticated modelling of the process is required. The presence of a weak wind in optical spectra (M\'endez et al. 1988) also casts doubt on the scenario described by Garc{\'{\i}}a-D{\'{\i}}az et al. (2008).

\section{A newly detected ring of candidate LIS}
\label{sec:lis}
Two groups of low-ionisation structures (LIS, Gon{\c c}alves et al. 2001) positioned outside NGC~1360 were analysed in the Goldman et al. (2004) and Garc{\'{\i}}a-D{\'{\i}}az et al. (2008) studies. They show approximately equal H$\alpha$ and [N~II] emission line intensities as expected for such structures (e.g. $-2\le$ log([NII]/H$\alpha$) $\le0$, Raga et al. 2008). They are moving faster than the nebula expansion velocity with a kinematic age of $\sim$5000 yr, suggesting they were formed after the nebula (Goldman et al. 2004). The two LIS groups are located near either end of the polar axis of the nebula and Garc{\'{\i}}a-D{\'{\i}}az et al. (2008) suggested their apparent high degree of collimation could be attributed to the since-disproven magnetic stellar wind. 

An additional isolated low-ionisation knot, not associated with either polar LIS group, was identified by Garc{\'{\i}}a-D{\'{\i}}az et al. (2008). The spectrum of the knot (figure 2 of Garc{\'{\i}}a-D{\'{\i}}az et al. 2008) shows approximately equal H$\alpha$ and [N~II] emission, and we measured a RV of 40 km s$^{-1}$ or $v_\mathrm{knot}=80^{+37}_{-18}$ km s$^{-1}$ deprojected for $i=30\pm10$ deg (Goldman et al. 2004). This knot, together with the increasing importance of LIS as a tracer of post-CE PNe (e.g. Miszalski et al. 2009b; Miszalski et al. 2011a, 2011c; Corradi et al. 2011; Boffin et al. 2012; Jones et al. 2014; Manick et al. 2015), prompted us to search for additional LIS using archival ESO Very Large Telescope (VLT) FORS2 images (Appenzeller et al. 1998) taken under programme ID 60.A-9800(C) on 8 Dec 2011. An H\_Alpha$+$83 image of 180 s (central wavelength $\lambda_c=656.3$ nm, width $W_0=7.3$ nm), an I\_BESS$+$77 image of 30 s ($\lambda_c=768.0$, $W_0=138.0$ nm) and an OIII$+$50 image of 180 s ($\lambda_c=500.1$ nm, $W_0=5.7$ nm) were reduced using the FORS pipeline with bias and sky flat frames. The polar LIS are located outside the field of view of these images.

\begin{figure*}
   \begin{center}
      \includegraphics[scale=0.60]{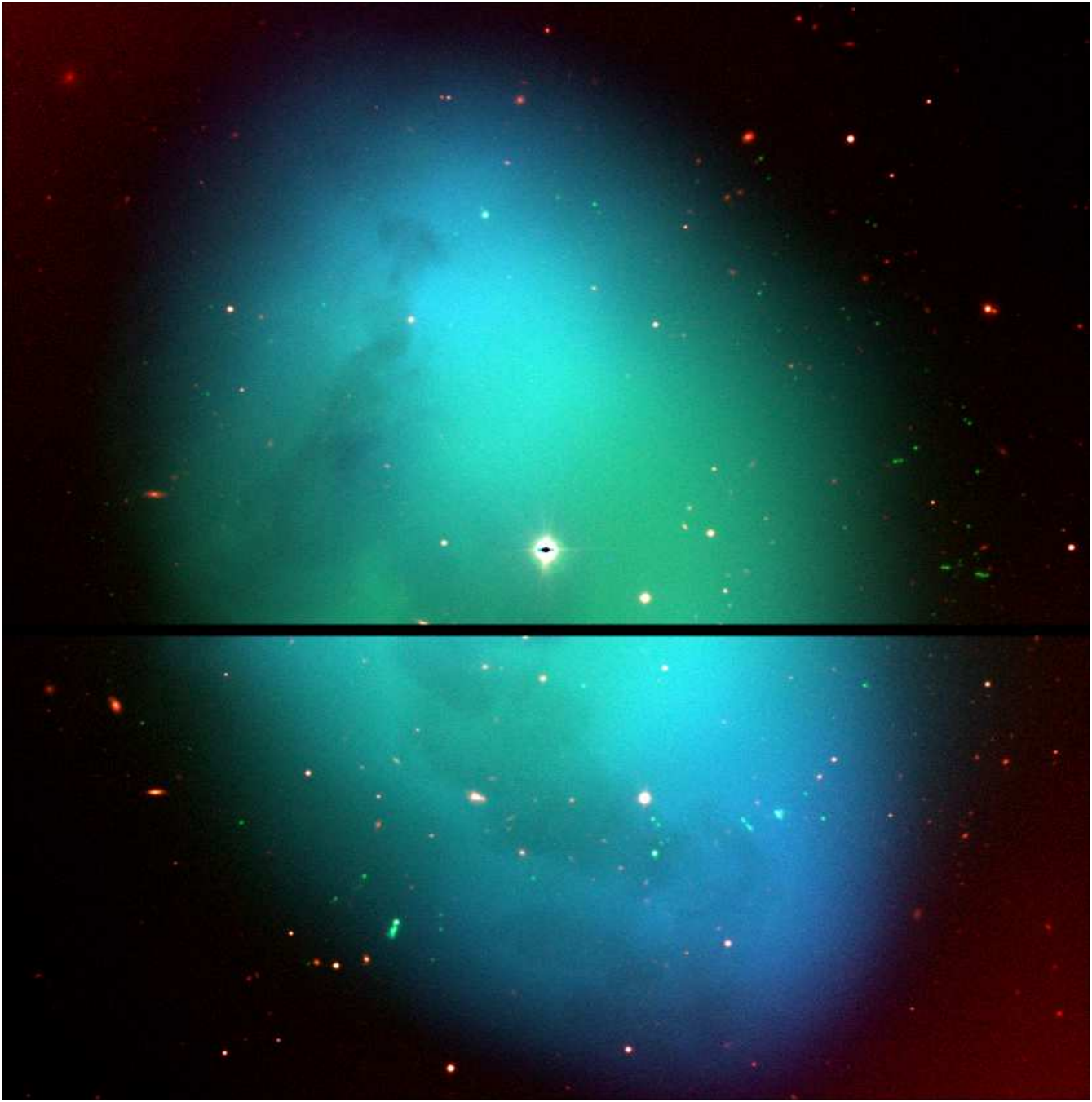}
   \end{center}
   \caption{Colour-composite image of NGC~1360 made from VLT FORS2 images taken with $I$-band (red), H$\alpha$+[N~II] (green) and [O~III] (blue) filters. Newly identified candidate low-ionisation structures (LIS) stand out as green sources, due to their H$\alpha$ and [N~II] emission, and form a ring around the central star (object just above the centre). Continuum objects (field stars and galaxies) are visisble as white or faint red sources. The image measures $7.0\times7.1$ arcmin$^2$ with North up and East to the left and includes a 4\arcsec\ wide chip gap.}
   \label{fig:lis}
\end{figure*}

We searched for candidate LIS by blinking together the H$\alpha$+[N~II], $I$-band and quotient (H$\alpha$+[N~II]/$I$-band) images with the \textsc{ds9} program (Smithsonian Astrophysical Observatory 2000; Joye \& Mandel 2003). Emission line sources were effectively isolated against the continuum sources (field stars and galaxies) in the quotient image. To help identify the faint knots against the variable nebula background the image contrast and bias were frequently modified during the search. The process was repeated with some smoothing applied to the images to detect the faintest knots. Figure \ref{fig:lis} displays a colour-composite image where many of the 120 candidate LIS can be identified as green sources due to their prominence in the H$\alpha$+[N~II] image. Appendix \ref{sec:appendix} gives a list of all the candidate LIS that form a ring around the central star. More knots were detected on the southern and western sides of the nebula. The asymmetry may be real, or it might be explained by the southwest end of the nebula being closer to us (Goldman et al. 2004), rather than incompleteness in our search methodology.  

\begin{figure}
   \begin{center}
      \includegraphics[scale=0.48]{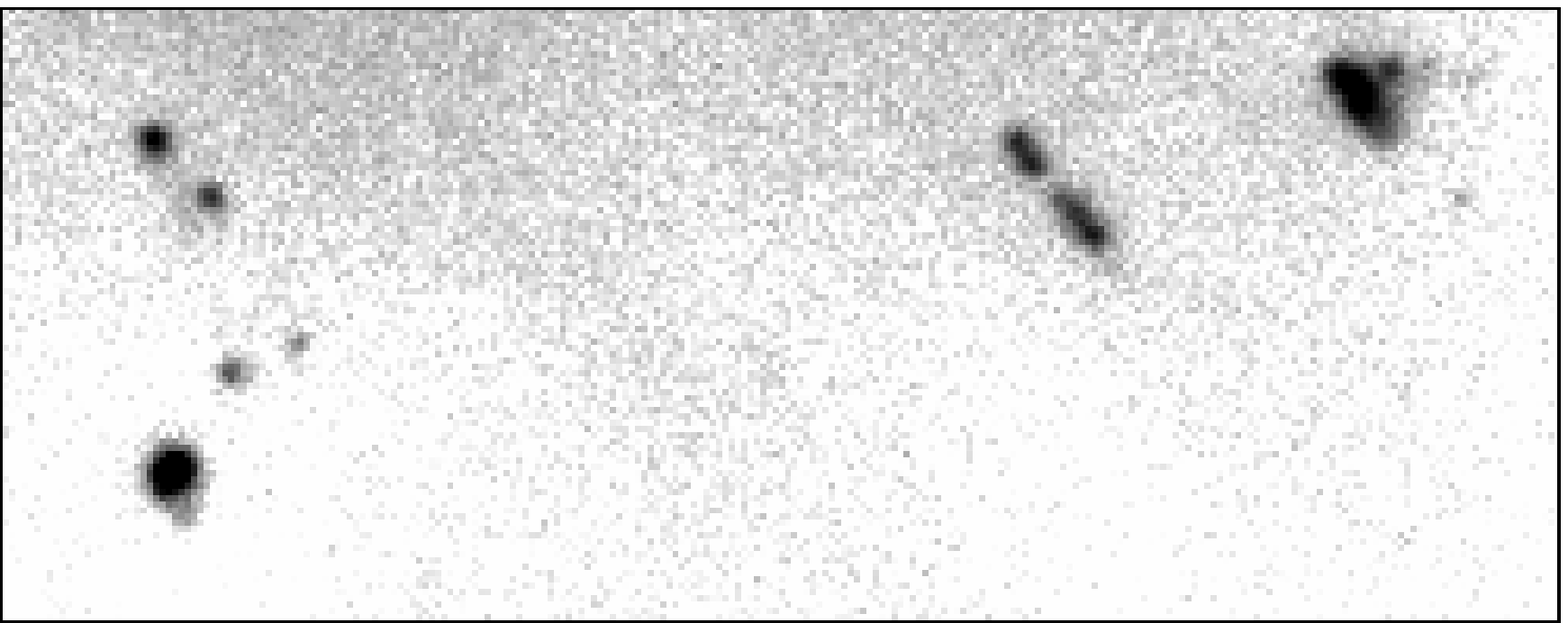}
   \end{center}
   \caption{Examples of elongated LIS candidates and some with photoevaporating tails in a 60$\times$24 arcsec$^2$ cutout from the H$\alpha$+[N~II] VLT FORS image. North is up and East to the left.}
   \label{fig:tails}
\end{figure}

\begin{figure}
   \begin{center}
      \includegraphics[scale=0.42]{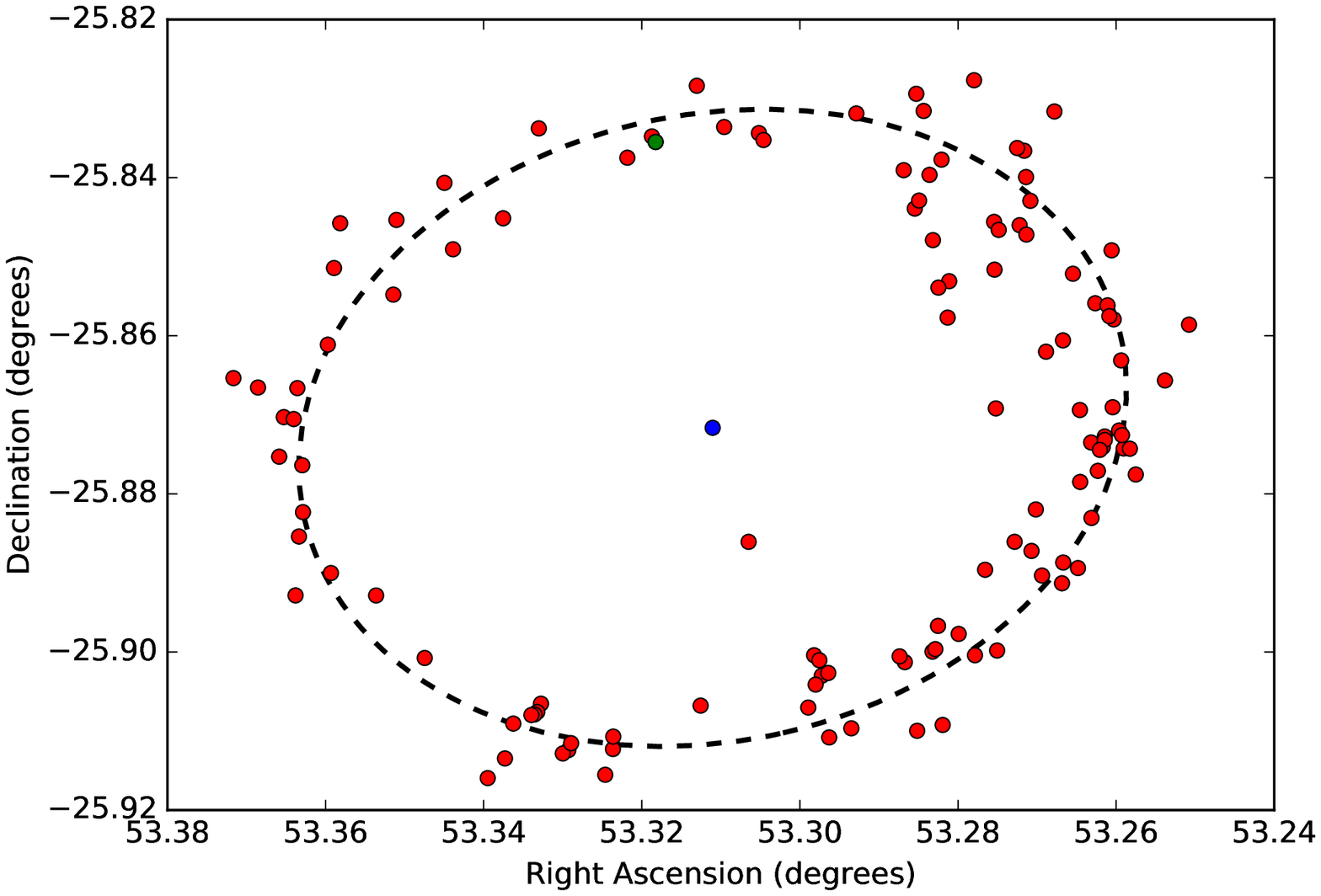}
   \end{center}
   \caption{Best-fit ellipse (dashed line) of the newly discovered candidate LIS (red dots) and the LIS knot (green dot) observed spectroscopically in figure 2 of Garc{\'{\i}}a-D{\'{\i}}az et al. (2008). The central star (blue dot) is the fixed centre of the ellipse. North is up and East to the left.}
   \label{fig:ringfit}
\end{figure}

Detailed spectroscopic and kinematic studies of the ring and its knots are beyond the scope of this paper. Based on the available observations we can, however, discuss some of its preliminary properties. Three observations suggest the knots are indeed LIS rather than some other phenomenon: (a) The approximately equal H$\alpha$ and [N~II] emission from a spectrum of one of the knots (Garc{\'{\i}}a-D{\'{\i}}az et al. 2008), (b) The strong morphological similarity with other rings of LIS (e.g. Corradi et al. 2011; Boffin et al. 2012), and (c) The appearance of photoevaporated tails in some knots (Figure \ref{fig:tails}) that point away from the central star and resemble the cometary globule variety of LIS (e.g. Matsuura et al. 2009). We used \textsc{kmpfit} of the Kapteyn Python Package (Terlouw \& Vogelaar 2016) to fit the ring with an ellipse fixed on the central star location. Figure \ref{fig:ringfit} displays the best fit ellipse which has parameters $a=171.3\pm13.6$\arcsec, $b=142.6\pm12.7$\arcsec\ and $\theta=-12.4\pm12.5$ deg. At $d=421$ pc the major and minor axes measure 0.35 and 0.29 pc, respectively. Assuming the ring is circular, the eccentricity of 0.55$^{+0.16}_{-0.38}$ corresponds to an inclination angle of $33.7^{+11.7}_{-23.7}$ deg in the plane of the sky. This is consistent with and supports the $i=30\pm10$ deg measured from nebula kinematics (Goldman et al. 2004). If the average motion of the ring is represented by the velocity of the knot from figure 2 of Garc{\'{\i}}a-D{\'{\i}}az et al. (2008) with $v_\mathrm{knot}=80^{+37}_{-18}$ km s$^{-1}$, then using $a=171.3\pm13.6$\arcsec we estimate a preliminary kinematic age of $4300\pm1300$ yrs for the ring. A more thorough kinematic study is required to distinguish whether the ring is coeval with the $\sim$5000 yr old polar LIS (Goldman et al. 2004) or possibly even the $\sim$7400 yr old main nebula, the age of the latter being recalculated using the best-fit parameters from Goldman et al. (2004) and $d=421$ pc.

\section{SALT HRS observations}
\label{sec:obs}
We used the High Resolution Spectrograph (HRS, Bramall et al. 2010, 2012; Crause et al. 2014) to observe HIP~16566. The HRS is a dual-beam, fibre-fed {\'e}chelle enclosed in a vacuum tank located in an insulated, temperature controlled enclosure in the spectrograph room of SALT. We selected the medium resolution (MR) mode of HRS that provides spectra covering 370--890 nm with resolving powers $R=\lambda/\Delta\lambda$ of 43000 and 40000 for the blue and red arms, respectively. During an observation a second fibre, separated at least 20\arcsec\ from the science fibre, simultaneously observes the sky spectrum. Both object and sky spectra are interleaved on the separate blue (2k$\times$4k) and red (4k$\times$4k) CCDs. Regular bias, ThAr arc and quartz lamp flat field calibrations are taken as part of SALT operations. 

Table \ref{tab:log} gives a list of the 14 HRS MR observations of HIP~16566. Initially we had set a minimum 25 day wait period before subsequent spectra were observed. In Jan 2017 we noticed variability from the first few observations of HIP~16566, prompting us to add more observations to the queue. These observations were made until HIP~16566 became inaccessible to SALT in early Mar 2017. Basic processing of the data was performed by \textsc{pysalt} (Crawford et al. 2010) before the pipeline developed by A.~Y. Kniazev (see Kniazev et al. 2016) reduced the data and distributed the data products. The pipeline is based on the \textsc{midas} packages \textsc{echelle} (Ballester 1992) and \textsc{feros} (Stahl, Kaufer \& Tubbesing 1999). An overview of the processing steps performed by the pipeline is given by Kniazev et al. (2016). We made use of only the blue spectra that were not sky subtracted. The order-merged spectra from the pipeline were converted to a logarithmic wavelength scale using \textsc{iraf} before adding the heliocentric correction using the \textsc{velset} task of the \textsc{rvsao} package (Kurtz \& Mink 1998).

\begin{table}
   \centering
   \caption{Log of SALT HRS observations. The Julian day marks the start of the exposure.}
   \label{tab:log}
   \begin{tabular}{llrr}
      \hline
      Julian day & Date & Exptime & Hel. cor. \\
          & (YYYY-MM-DD) & (s)  & (km s$^{-1}$)\\
      \hline
      2457675.41096 & 2016-10-13 & 1500 & 7.8411  \\
      2457708.33082 & 2016-11-15 & 400  &$-$4.3685 \\
      2457734.49451 & 2016-12-11 & 400  &$-$13.8206 \\
      2457766.37974 & 2017-01-12 & 400  &$-$20.8047 \\
      2457768.37785 & 2017-01-14 & 400  &$-$21.0544 \\
      2457792.31382 & 2017-02-07 & 415  &$-$21.8845 \\
      2457798.30426 & 2017-02-13 & 600  &$-$21.5027 \\
      2457800.31159 & 2017-02-15 & 440  &$-$21.3470 \\
      2457806.27403 & 2017-02-21 & 500  &$-$20.6100 \\
      2457807.28034 & 2017-02-22 & 440  &$-$20.4896 \\
      2457812.28064 & 2017-02-27 & 415  &$-$19.7269 \\
      2457815.28142 & 2017-03-02 & 200  &$-$19.1939 \\
      2457817.25951 & 2017-03-04 & 500  &$-$18.7689 \\
      2457820.25700 & 2017-03-07 & 440  &$-$18.1481 \\
      \hline
   \end{tabular}
\end{table}

\section{Analysis of HIP~16566}
\label{sec:analysis}
\subsection{Orbital parameters}
\begin{figure*}
   \begin{center}
\includegraphics[scale=0.17]{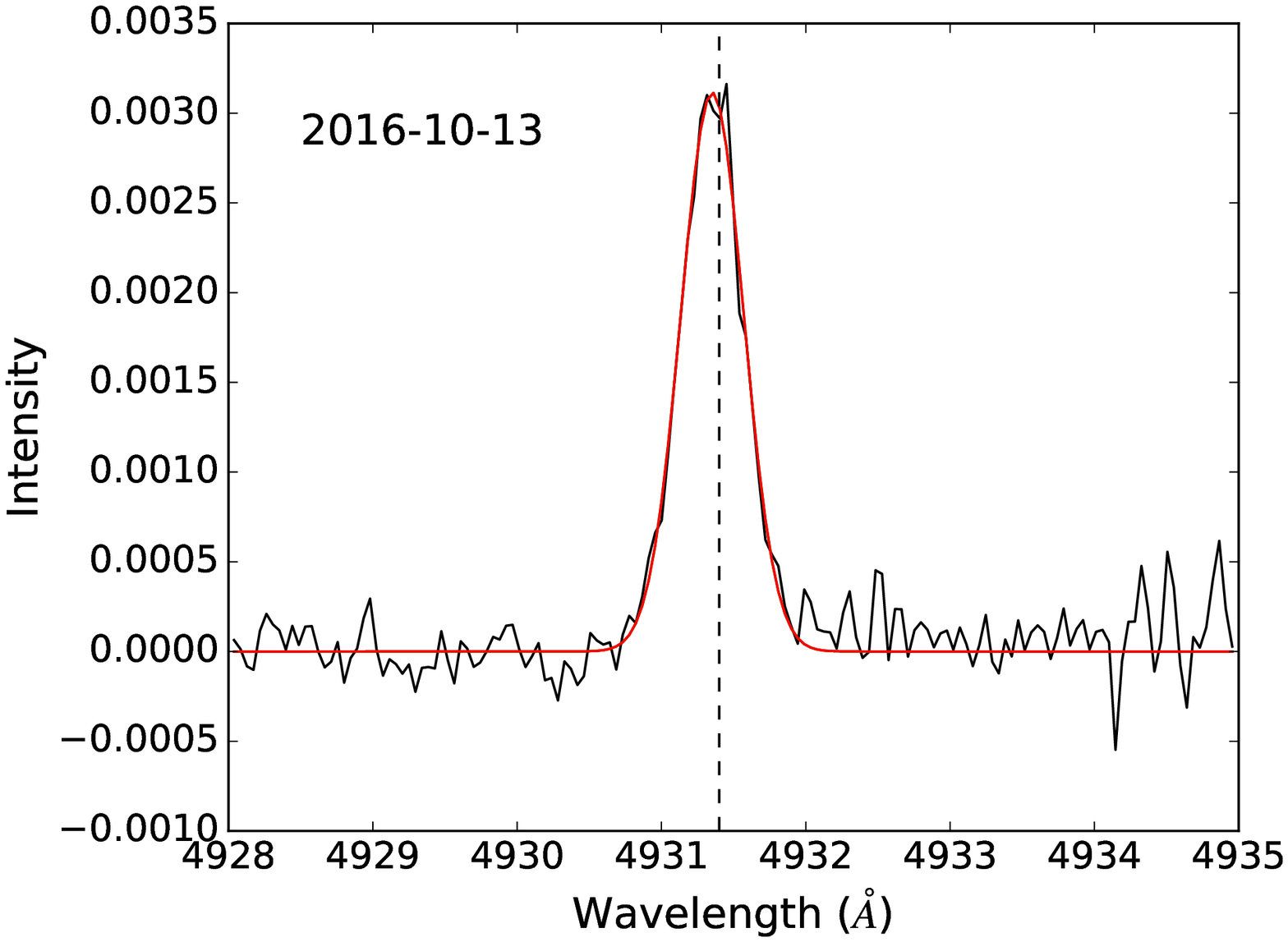}
\includegraphics[scale=0.17]{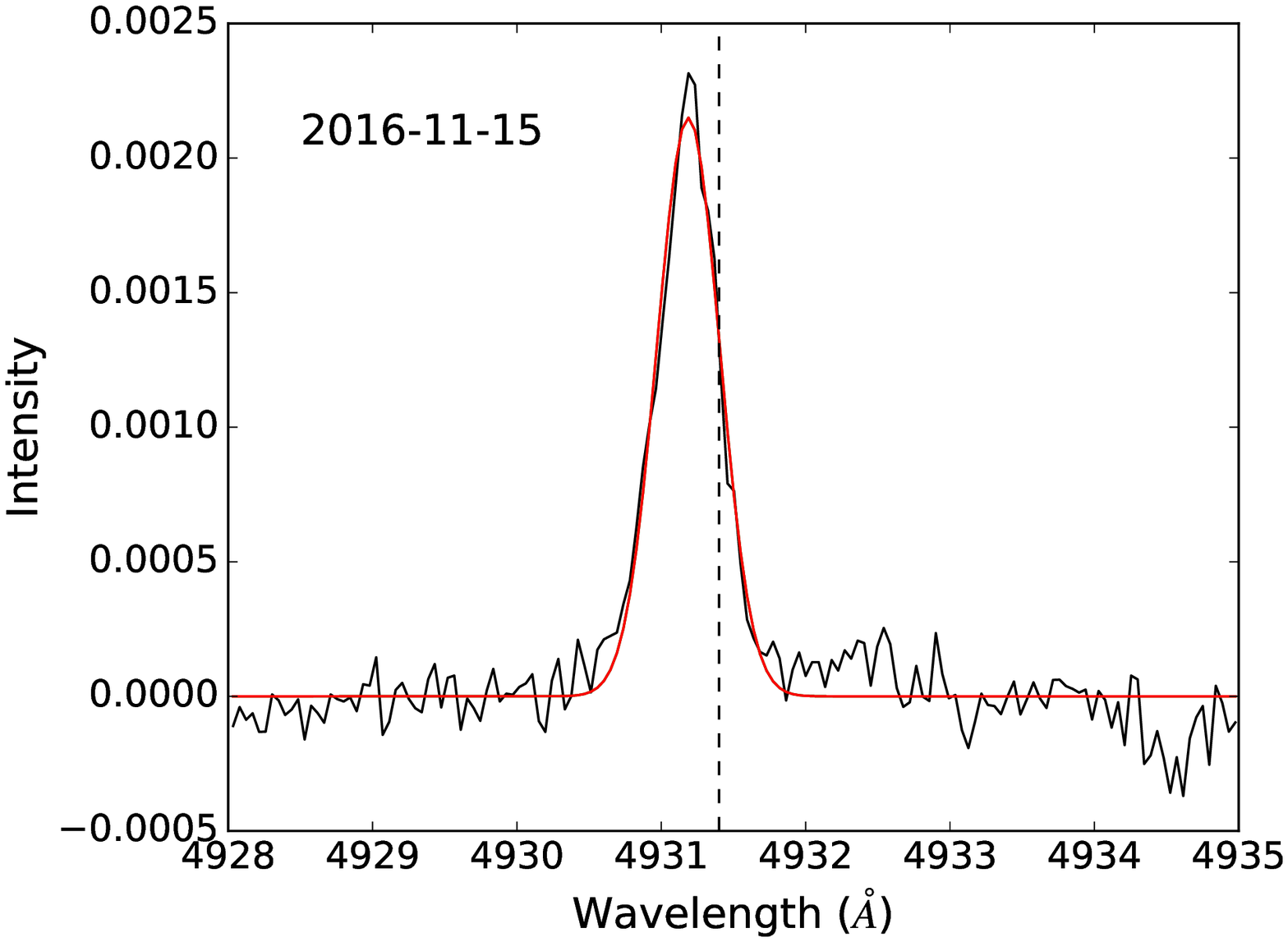}
\includegraphics[scale=0.17]{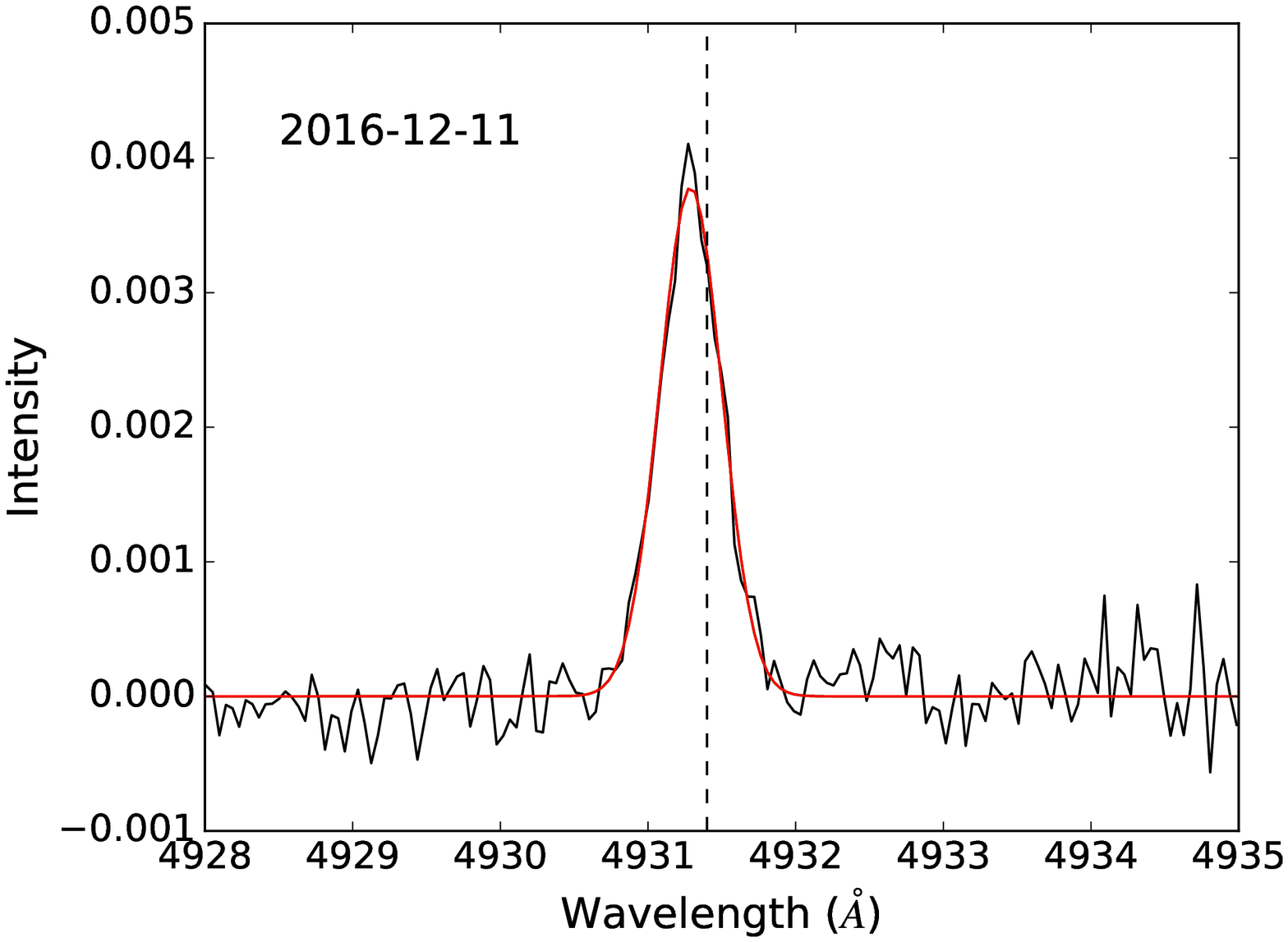}
\includegraphics[scale=0.17]{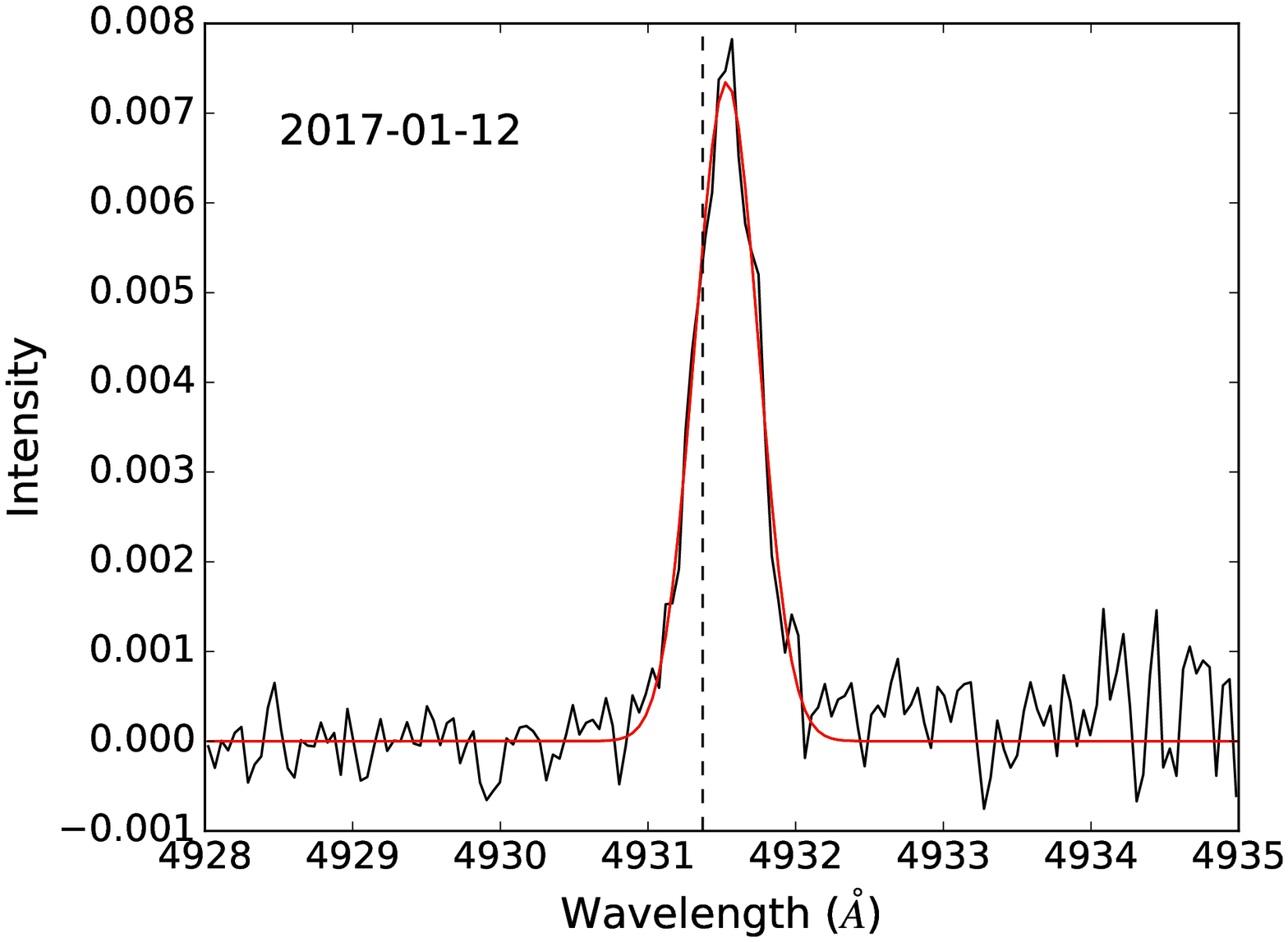}
\includegraphics[scale=0.17]{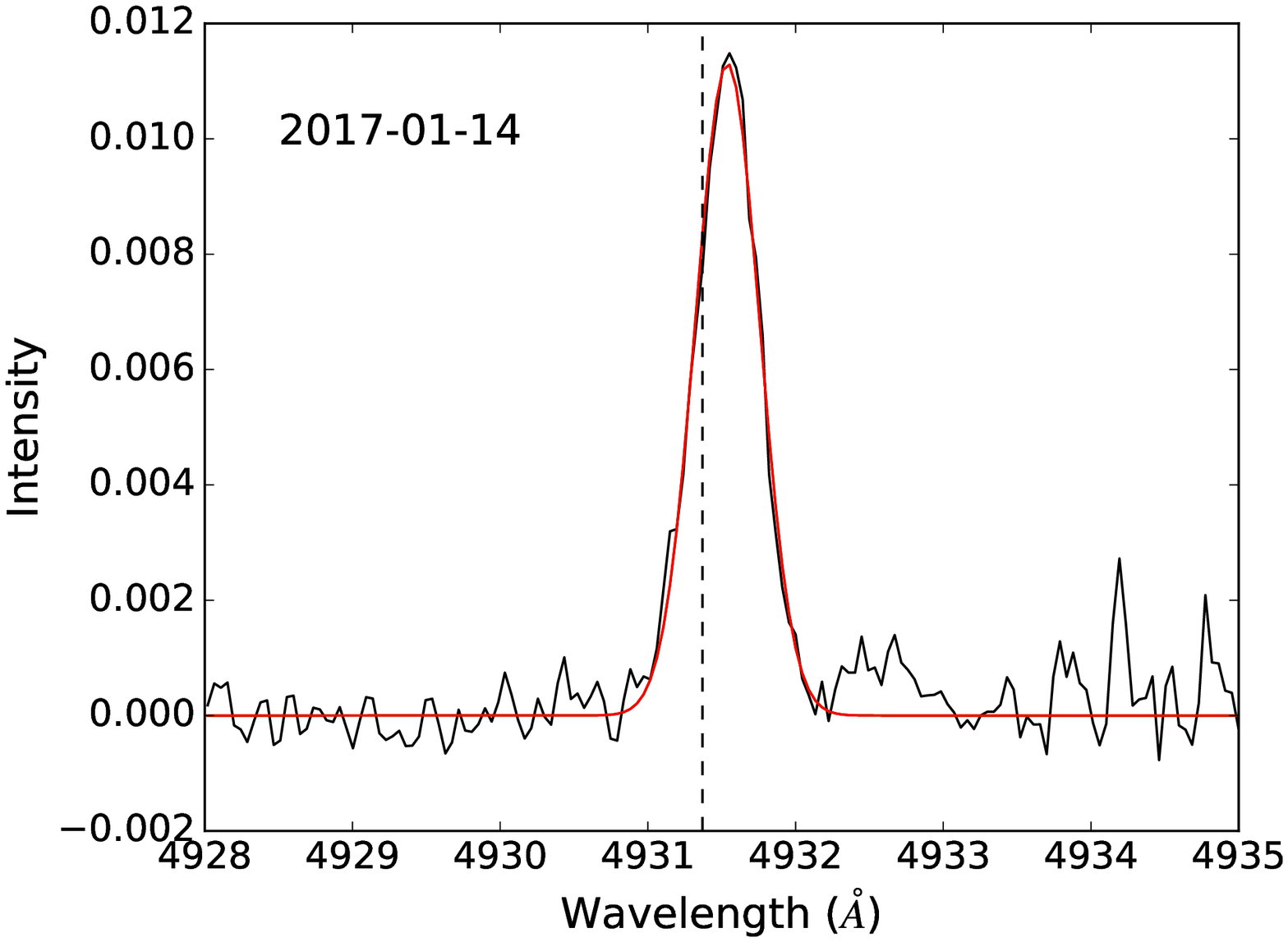}
\includegraphics[scale=0.17]{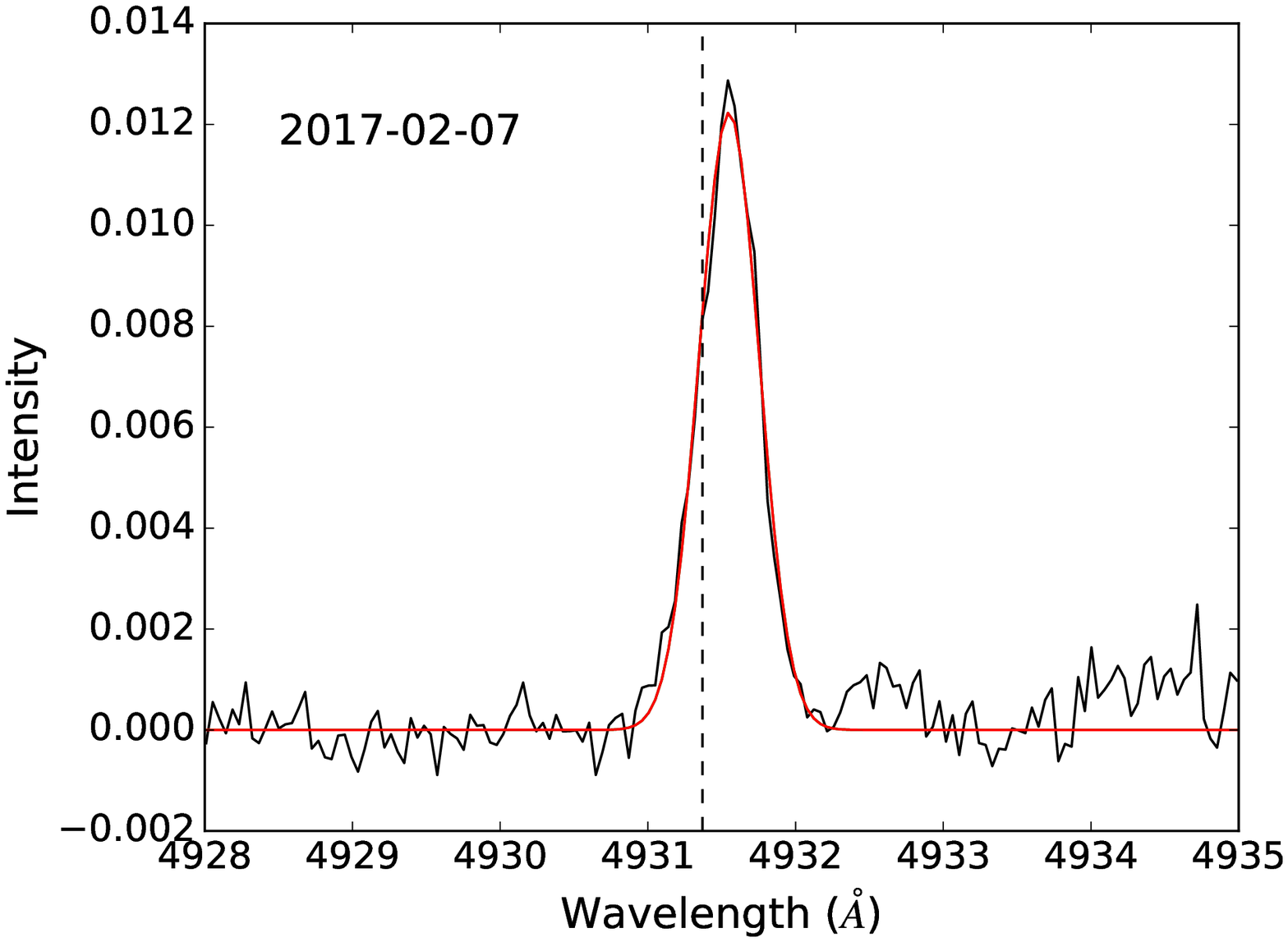}
\includegraphics[scale=0.17]{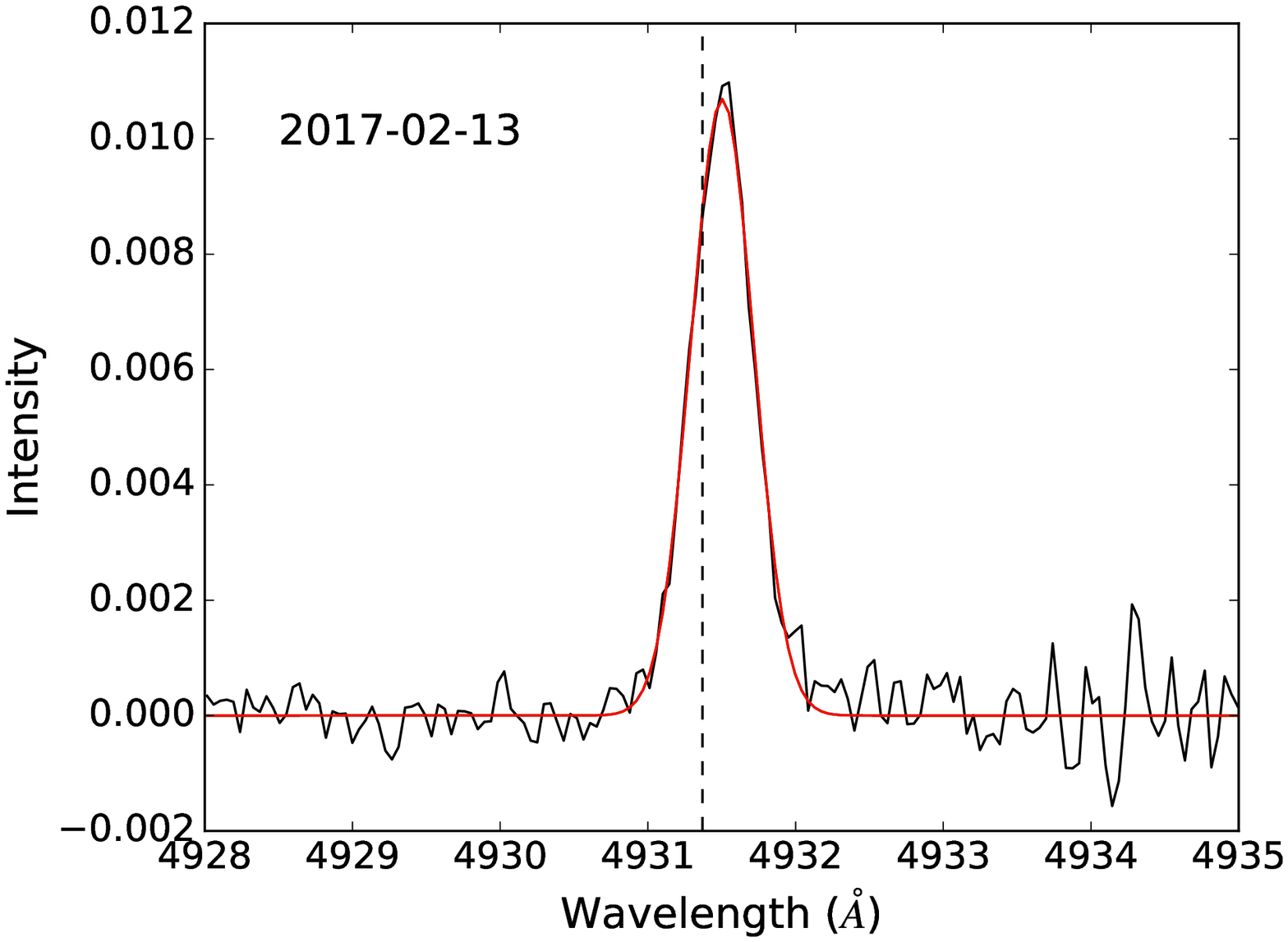}
\includegraphics[scale=0.17]{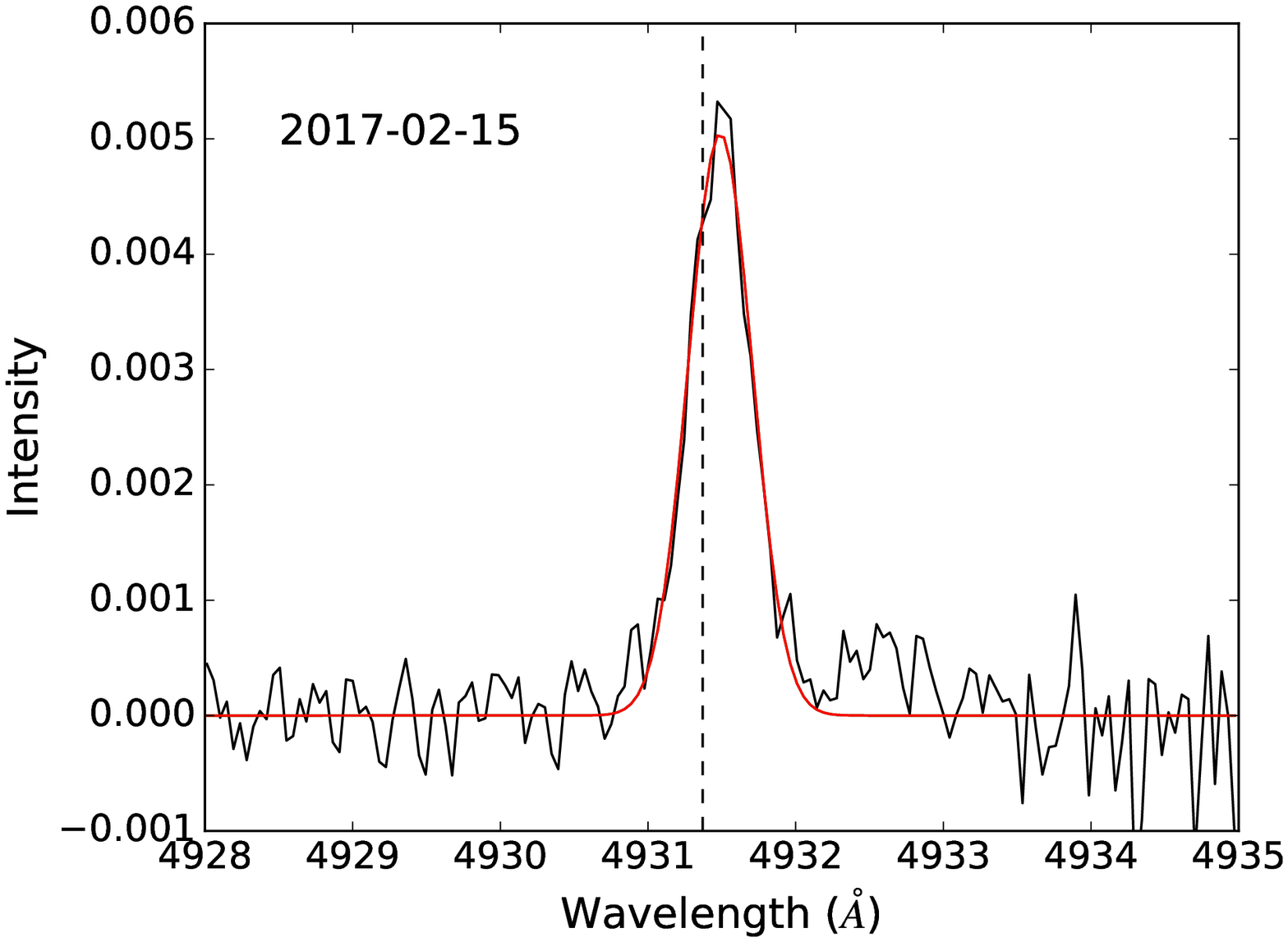}
\includegraphics[scale=0.17]{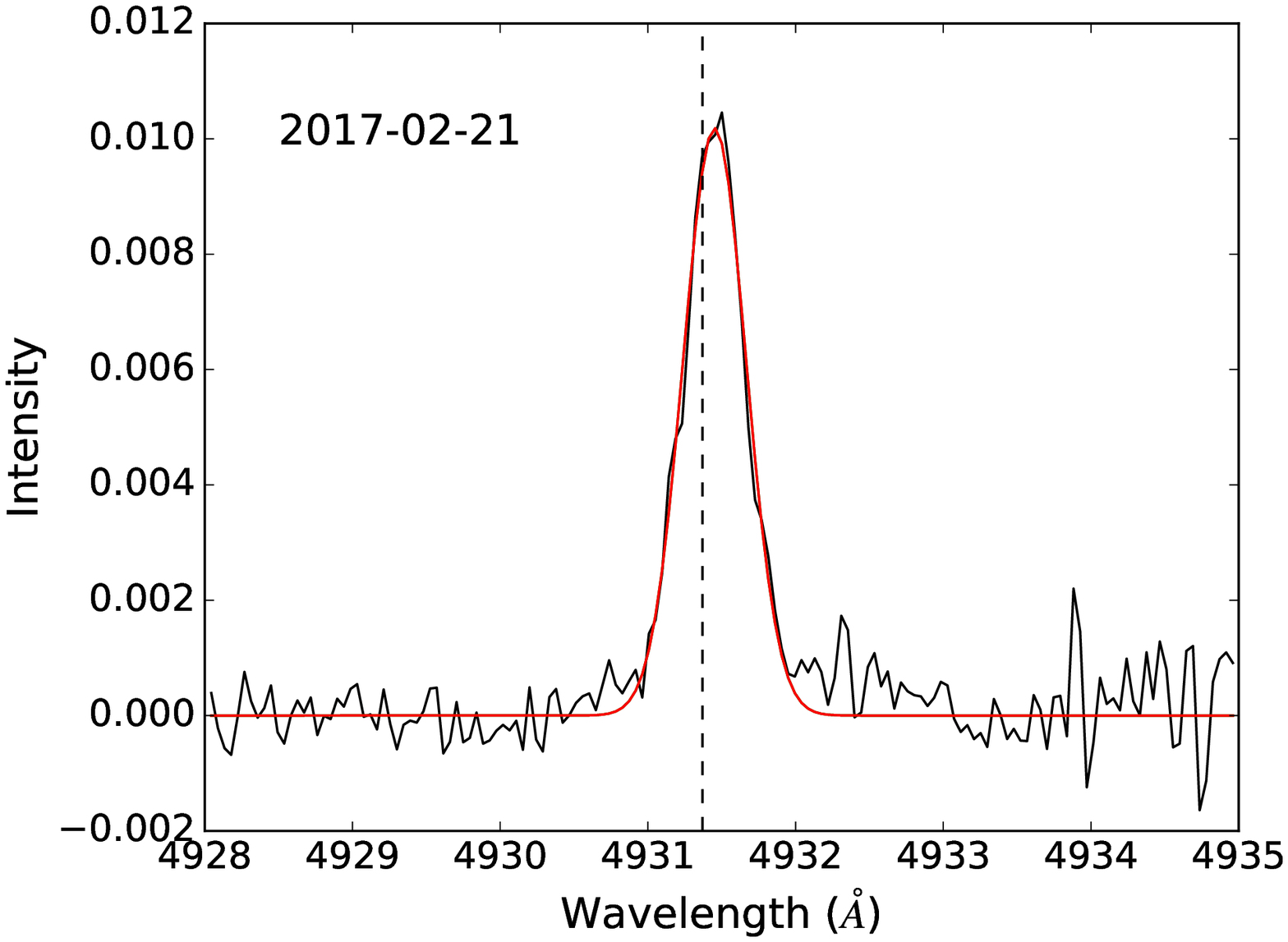}
\includegraphics[scale=0.17]{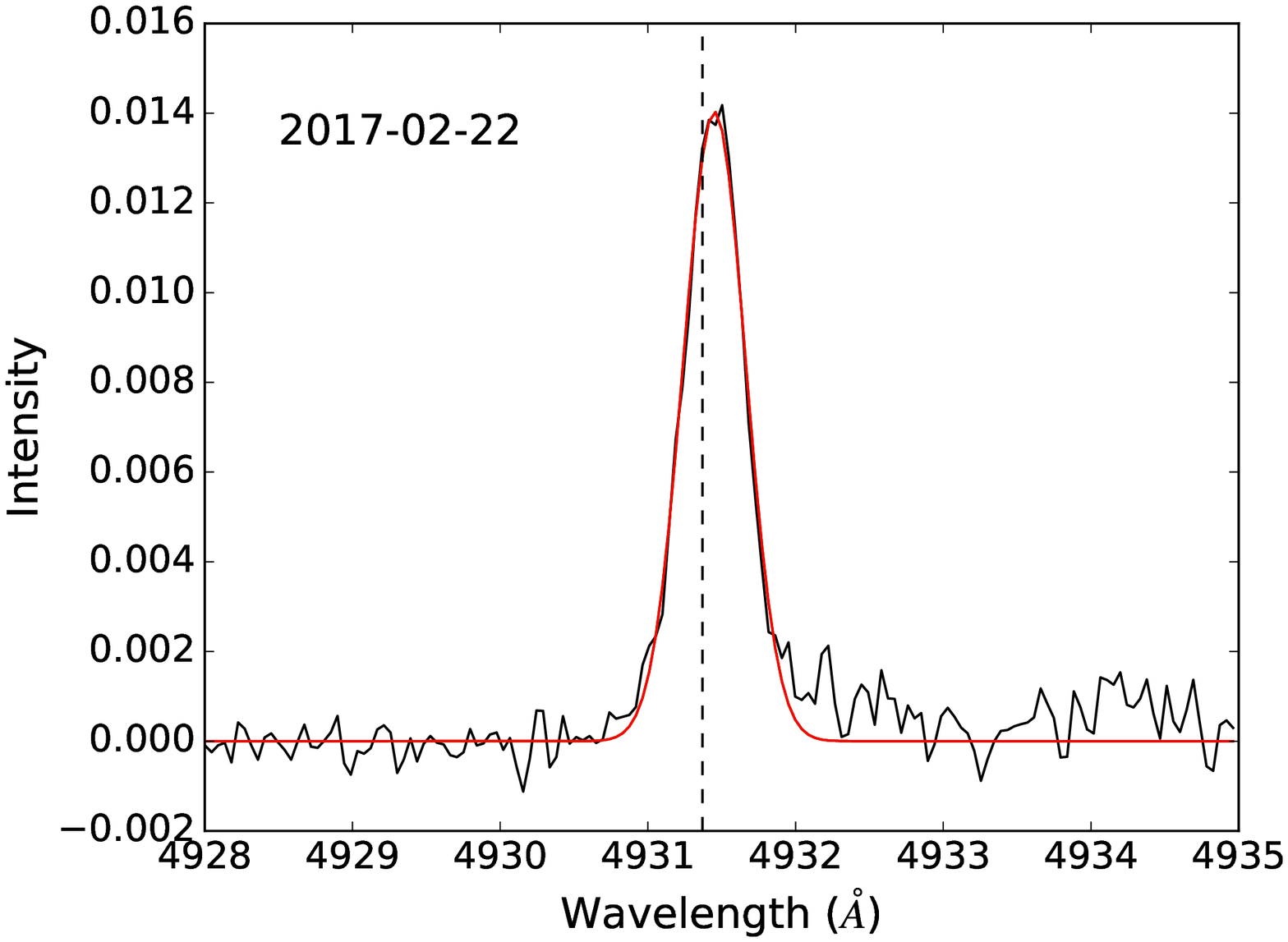}
\includegraphics[scale=0.17]{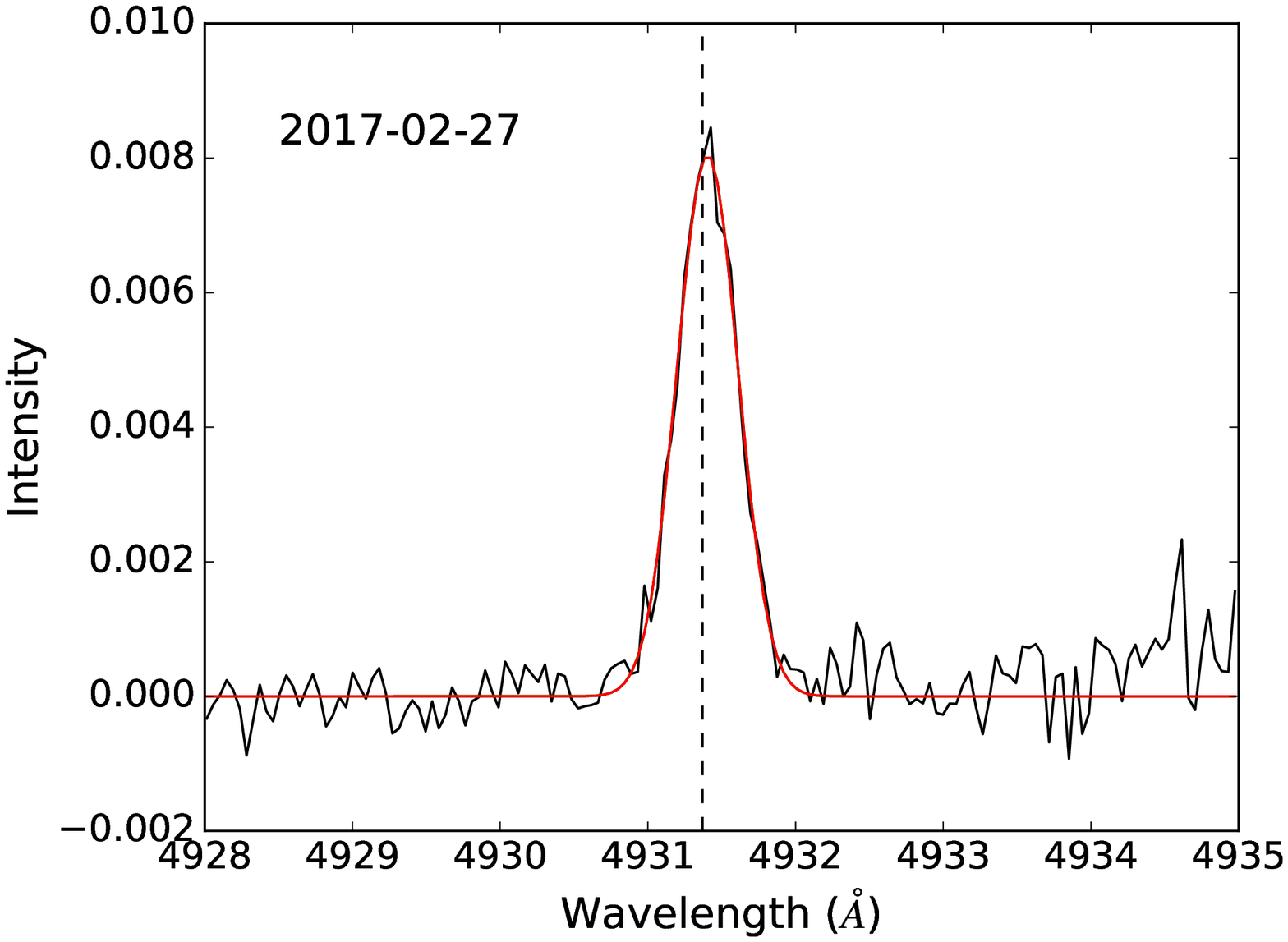}
\includegraphics[scale=0.17]{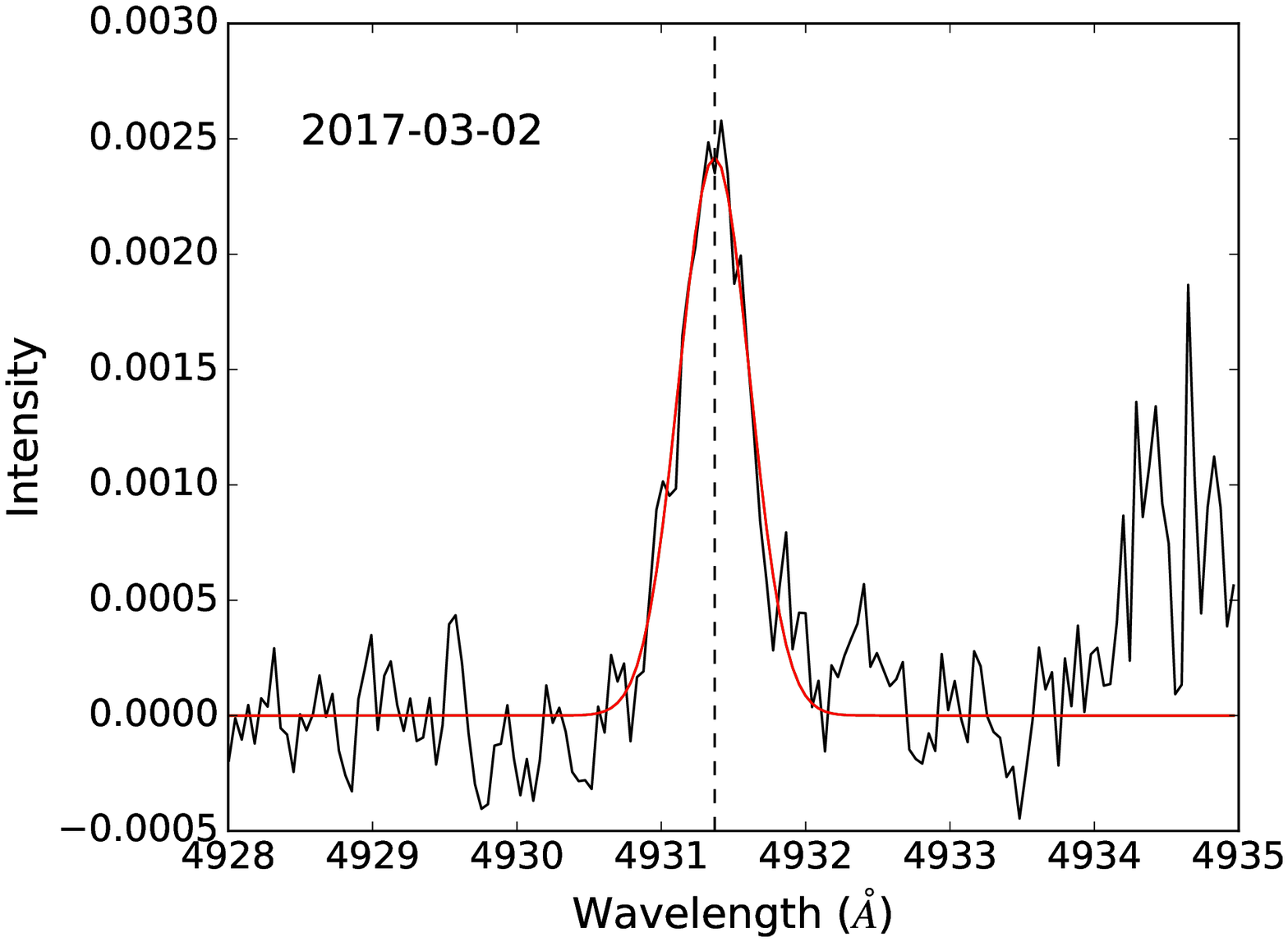}
\includegraphics[scale=0.17]{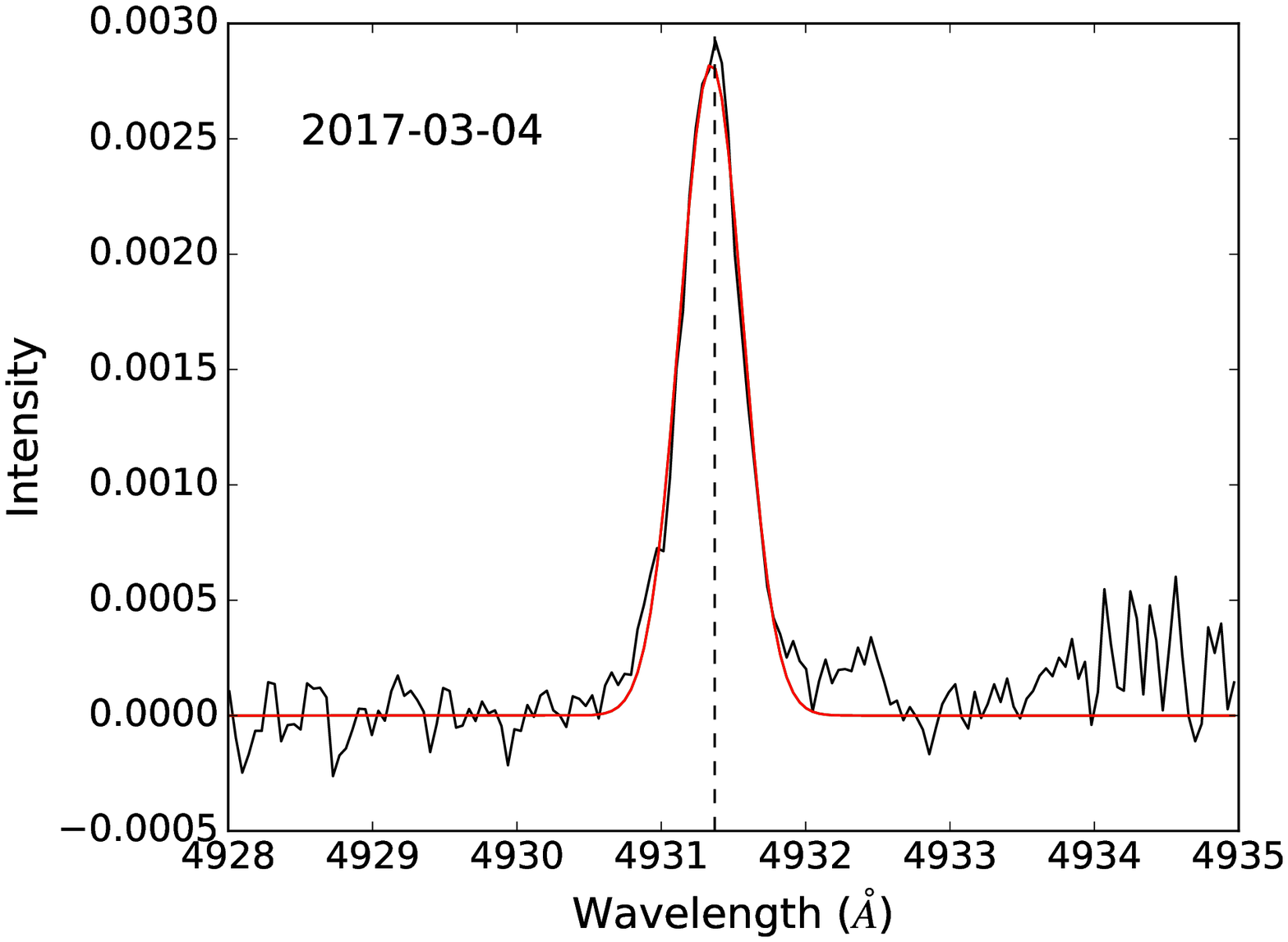}
\includegraphics[scale=0.17]{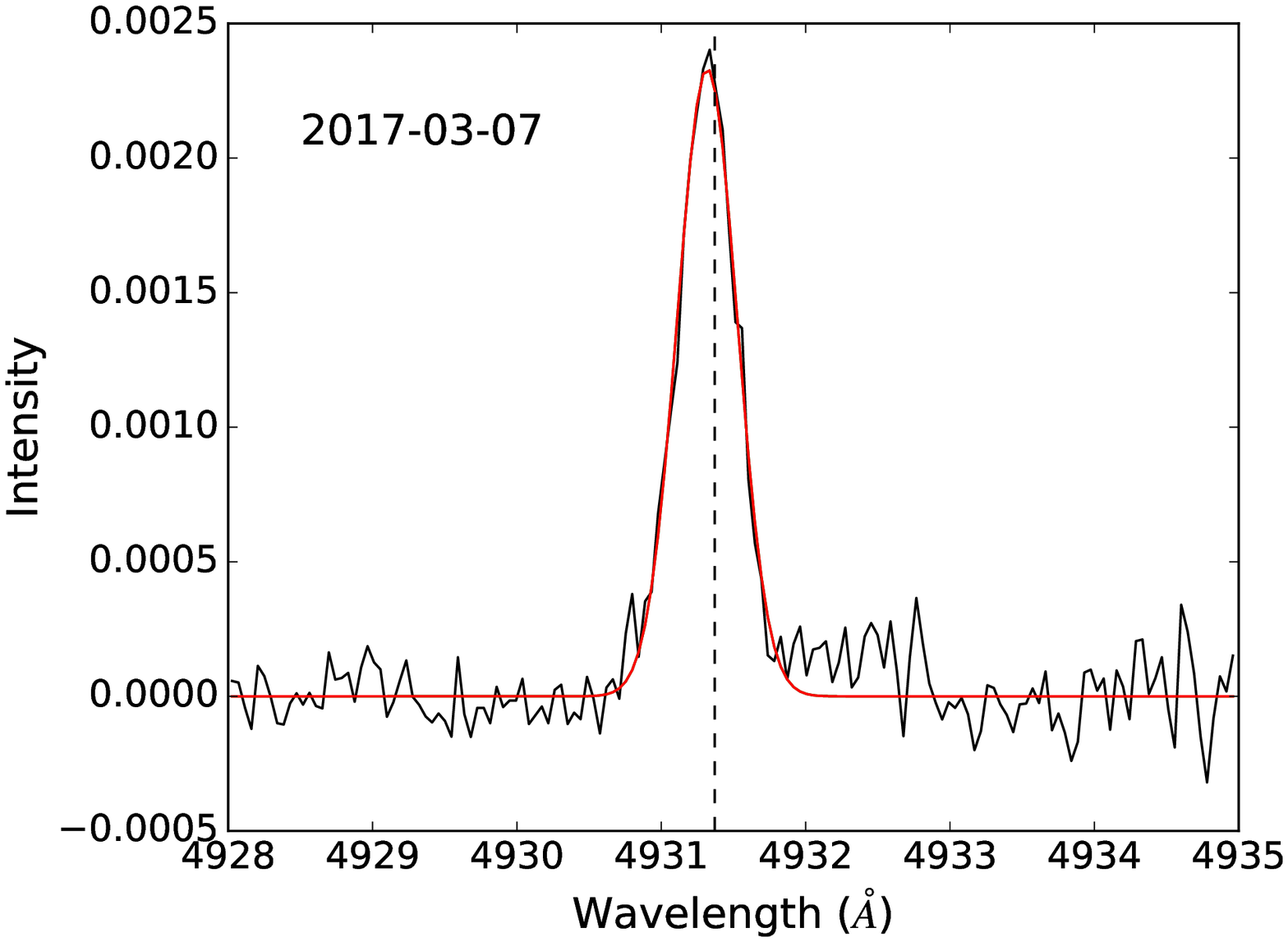}
   \end{center}
   \caption{Gaussian fits (red) to the O~V 4930 emission line of each spectrum (black). The dashed line corresponds to the expected position of O~V at $\gamma=67.09$ km s$^{-1}$ (Tab. \ref{tab:orbit}).}
   \label{fig:ovfits}
\end{figure*}

\begin{figure*}
   \begin{center}
      \includegraphics[scale=0.7,angle=270]{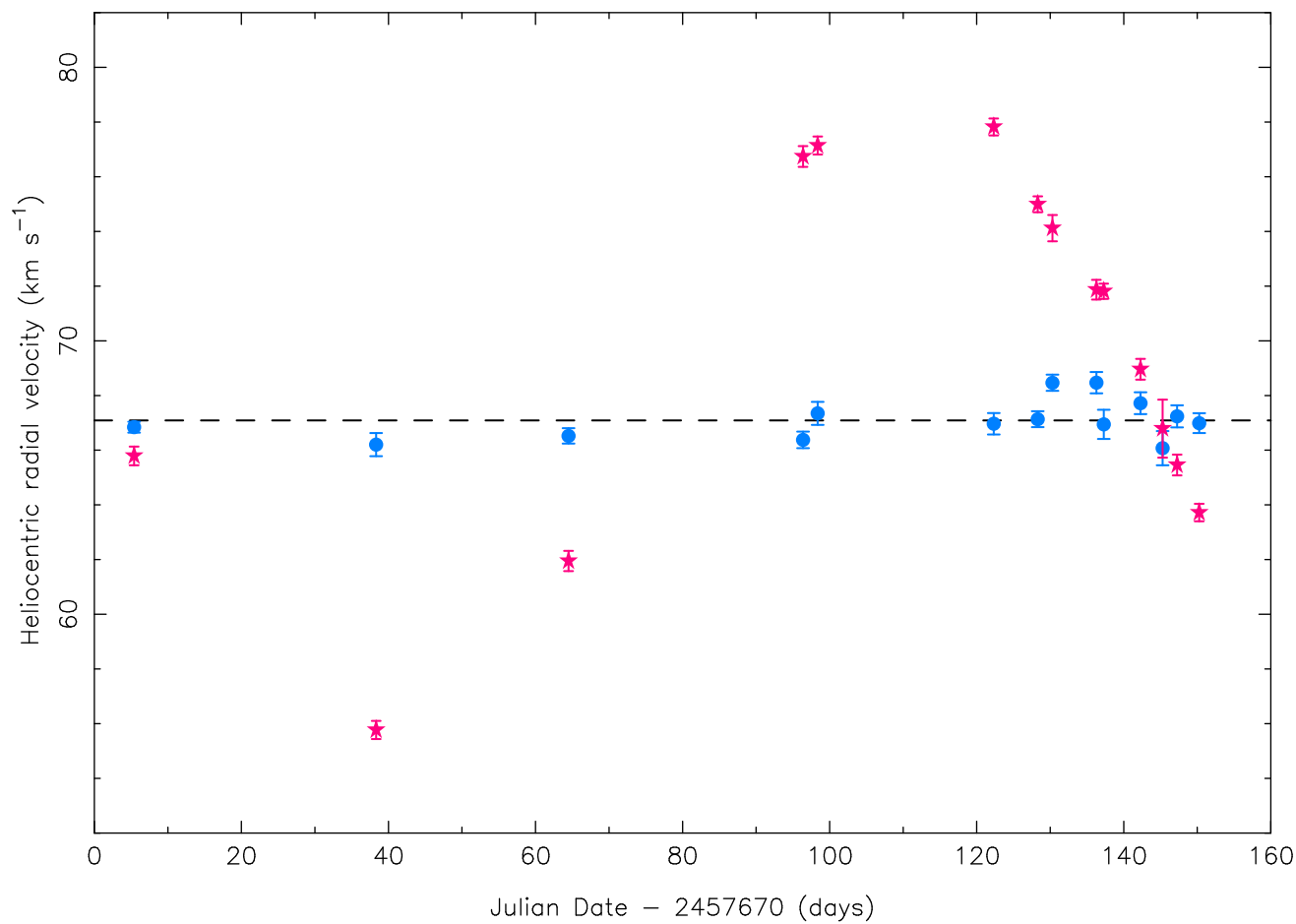}
   \end{center}
   \caption{SALT HRS RV time series of HIP 16566 measured from O~V $\lambda$4930 (pink stars). The dashed line indicates the average velocity of the fit in Tab. \ref{tab:orbit} and the blue dots are the [O~III] $\lambda$5007 nebula velocities shifted by $+$16.62 km s$^{-1}$ such that their average matches the dashed line.} 
   \label{fig:rvs}
\end{figure*}

\begin{figure*}
   \begin{center}
\includegraphics[scale=0.17]{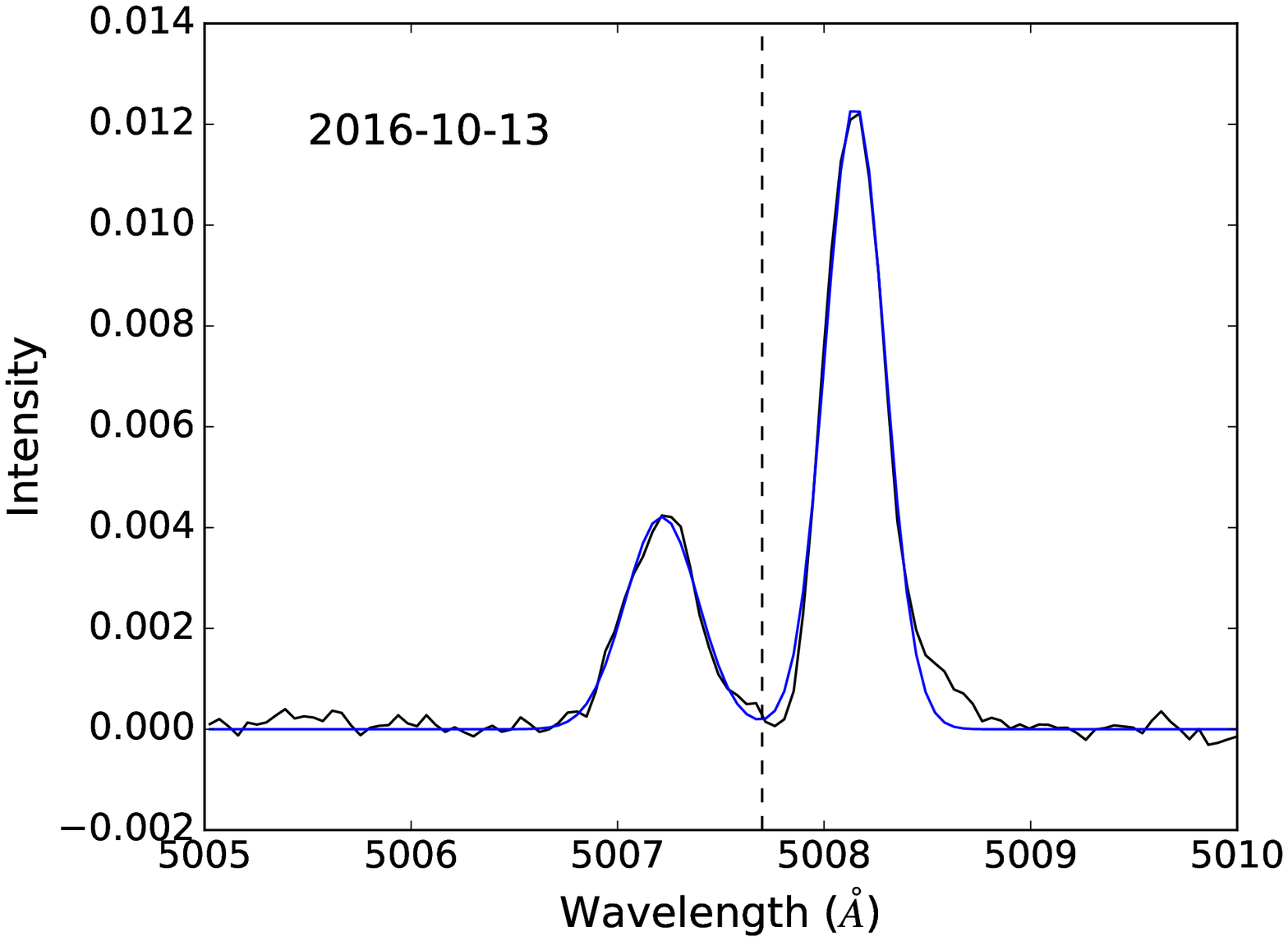}
\includegraphics[scale=0.17]{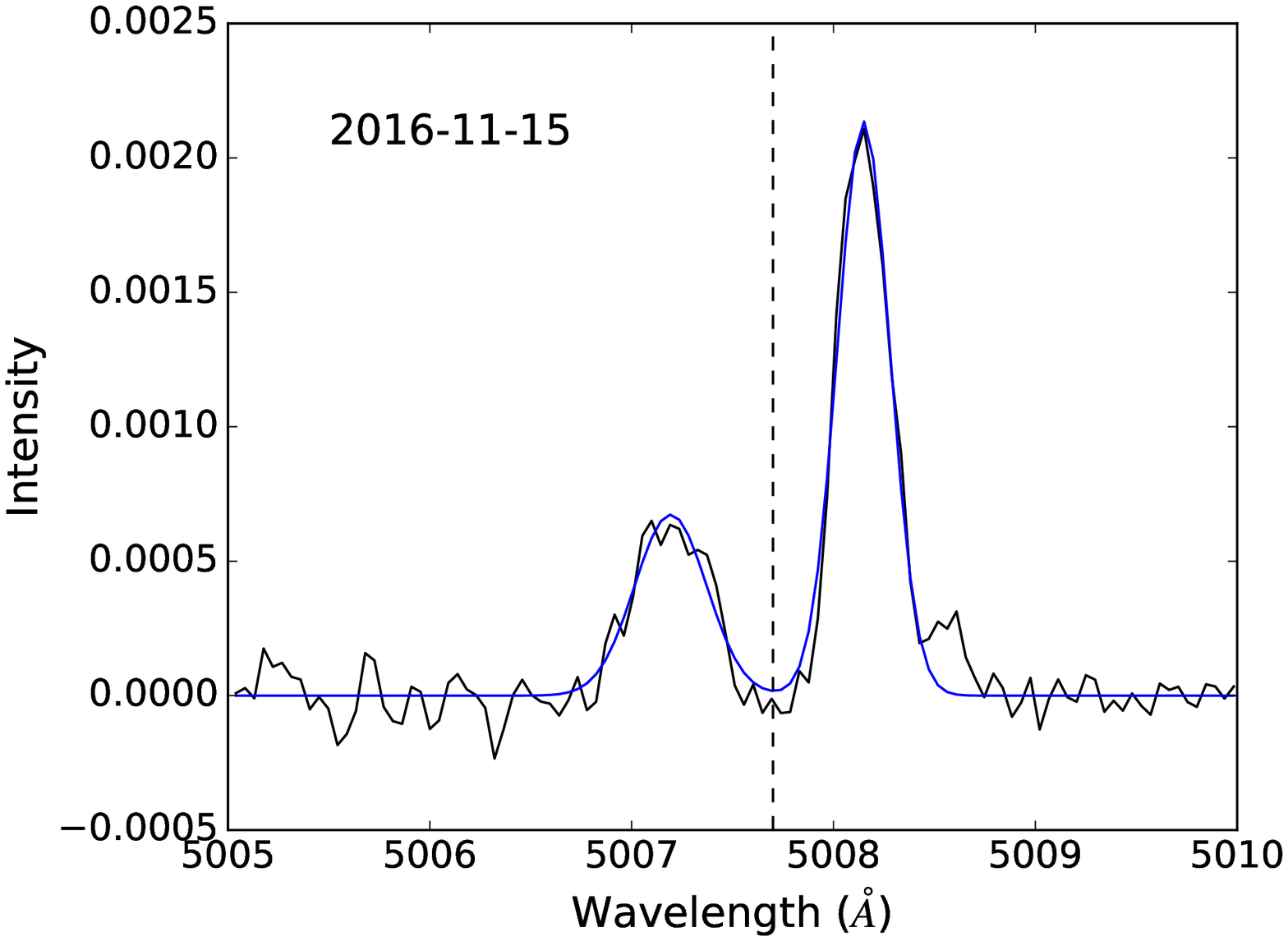}
\includegraphics[scale=0.17]{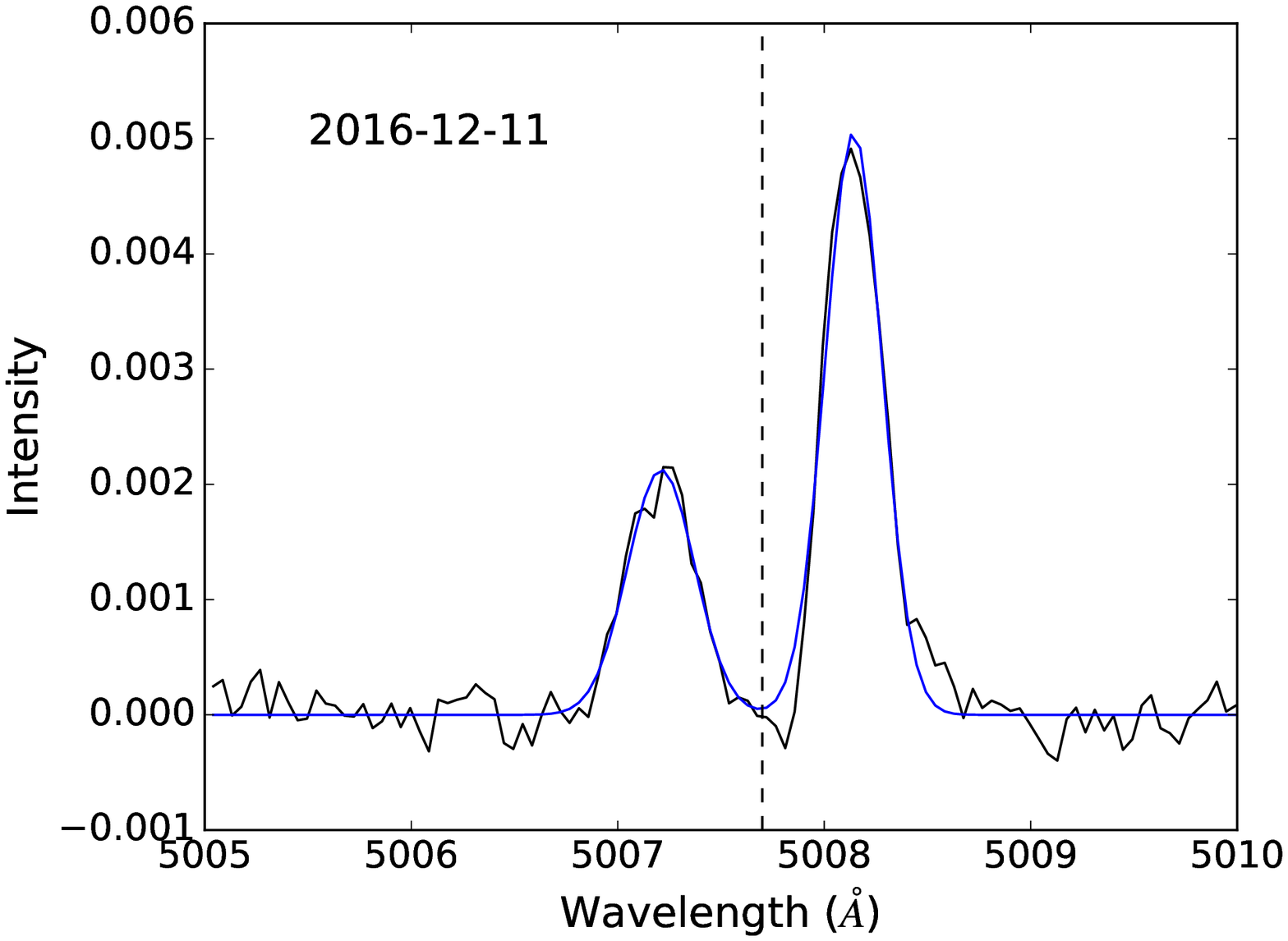}
\includegraphics[scale=0.17]{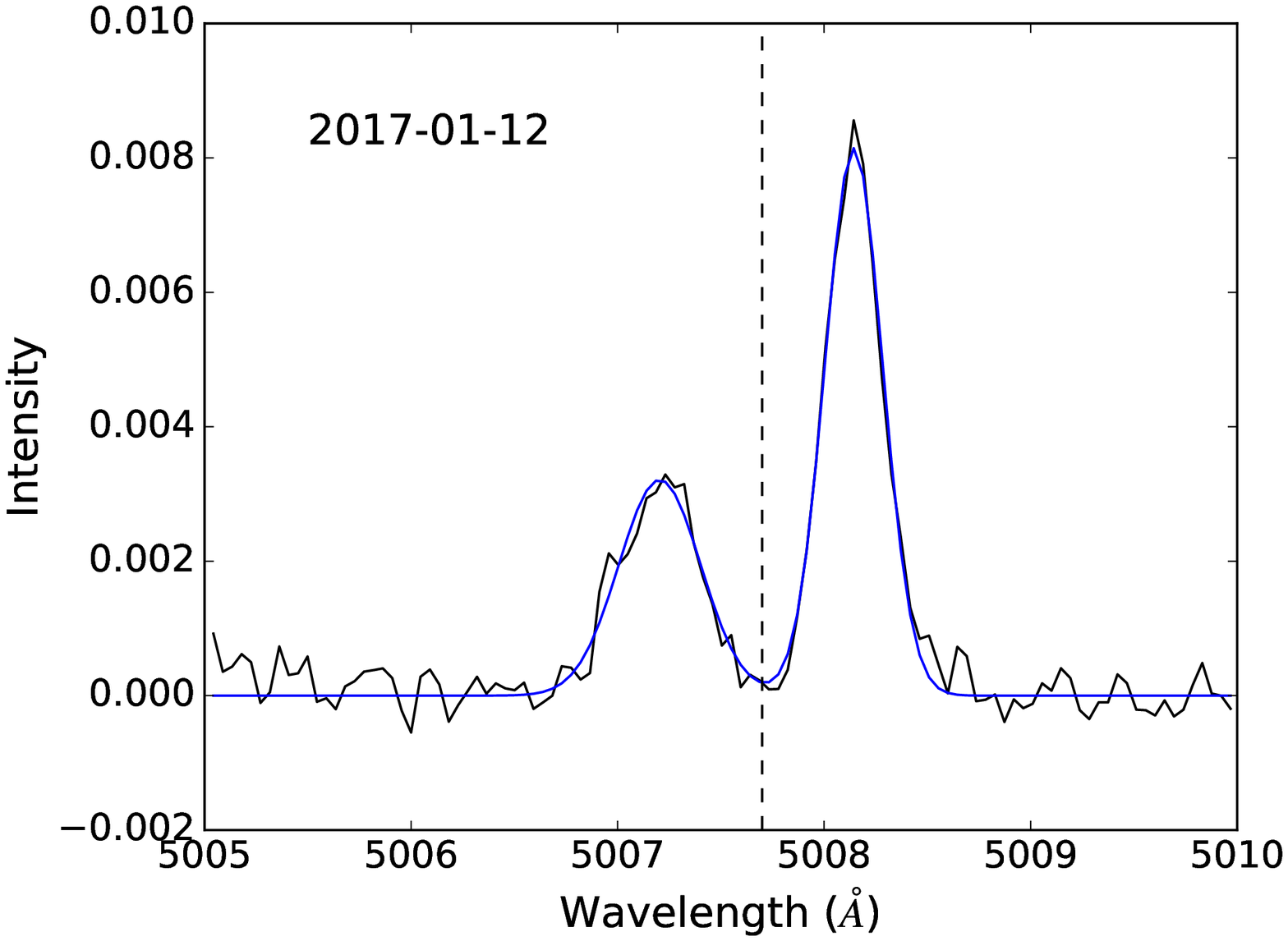}
\includegraphics[scale=0.17]{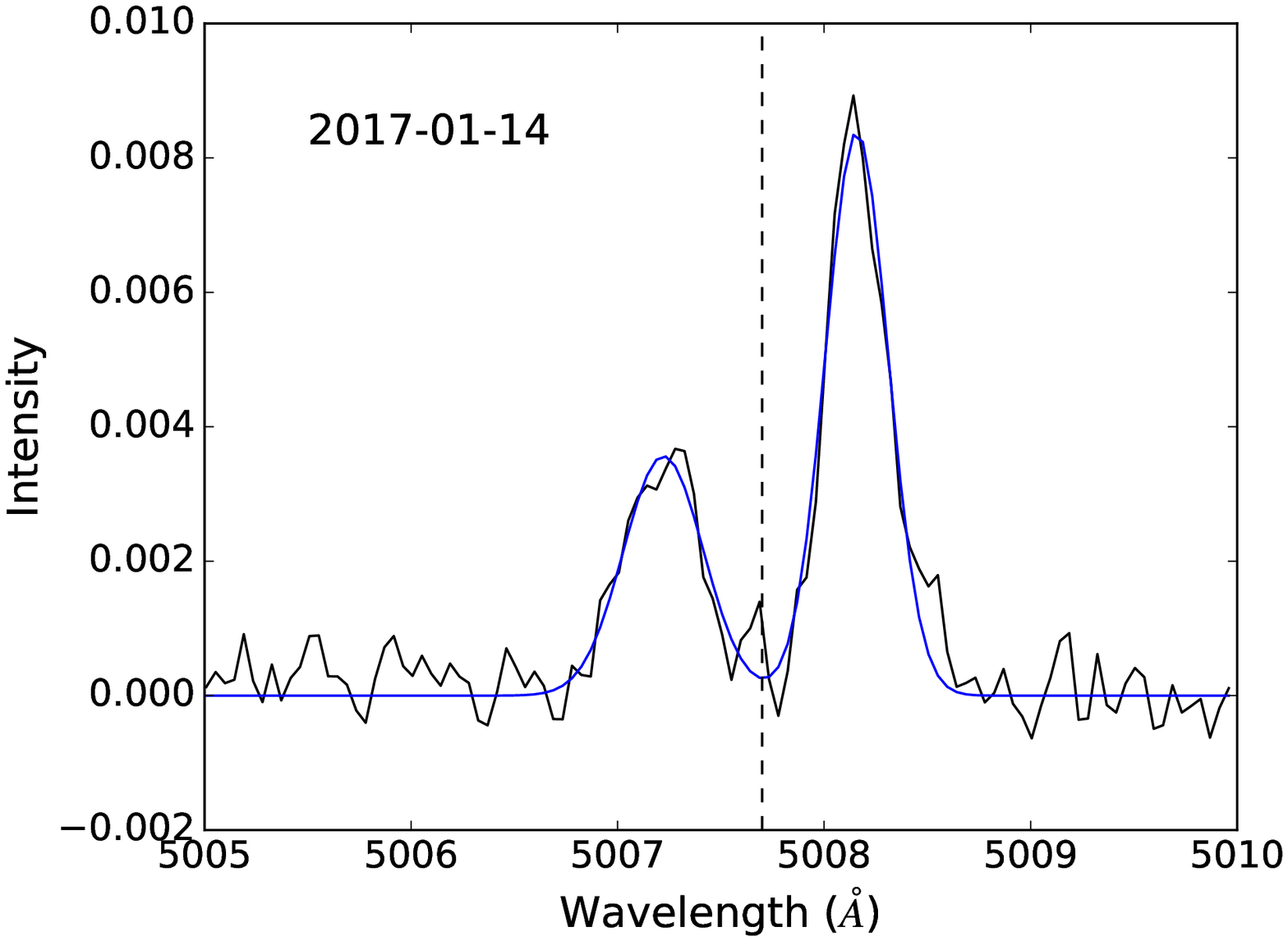}
\includegraphics[scale=0.17]{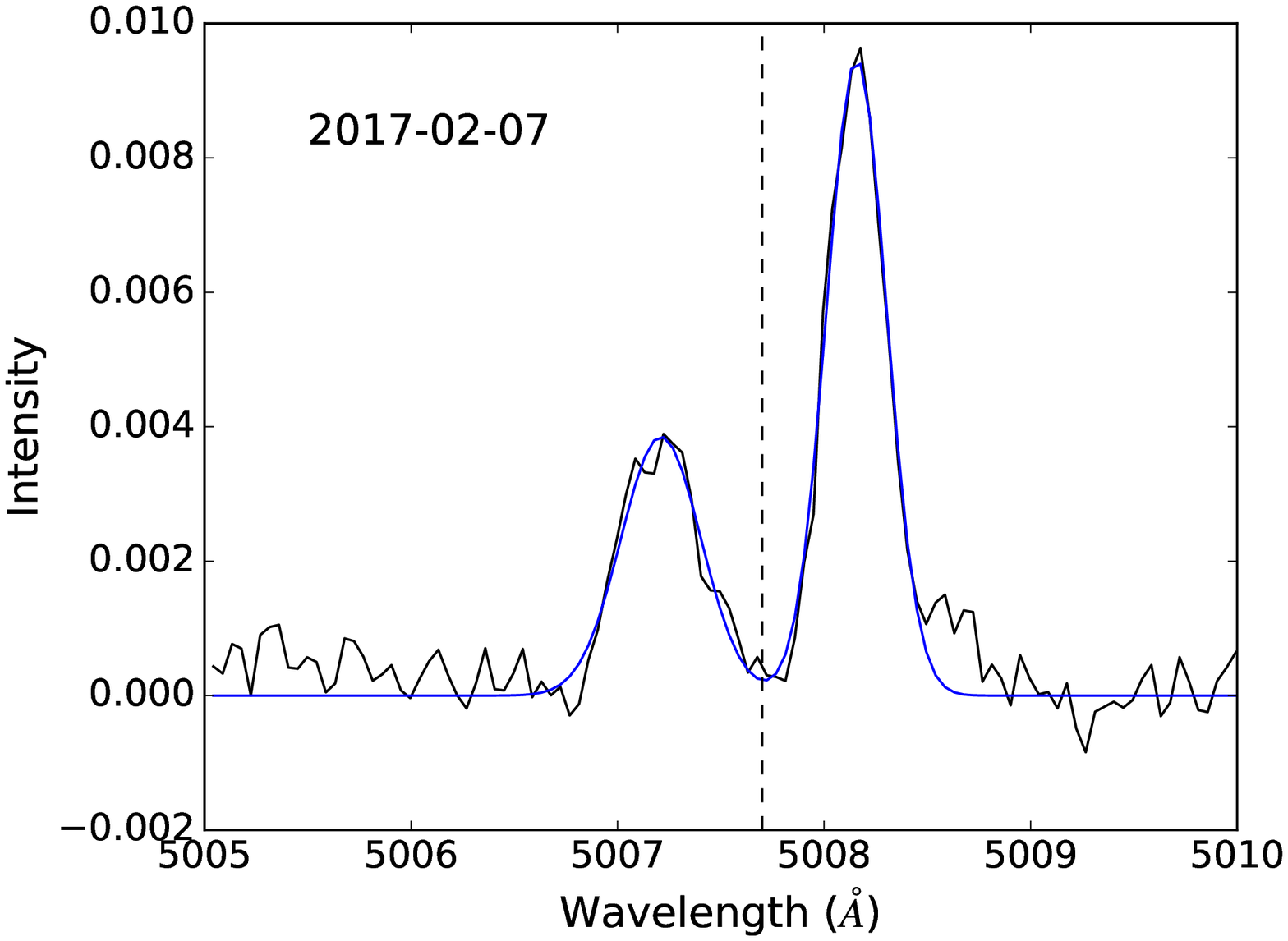}
\includegraphics[scale=0.17]{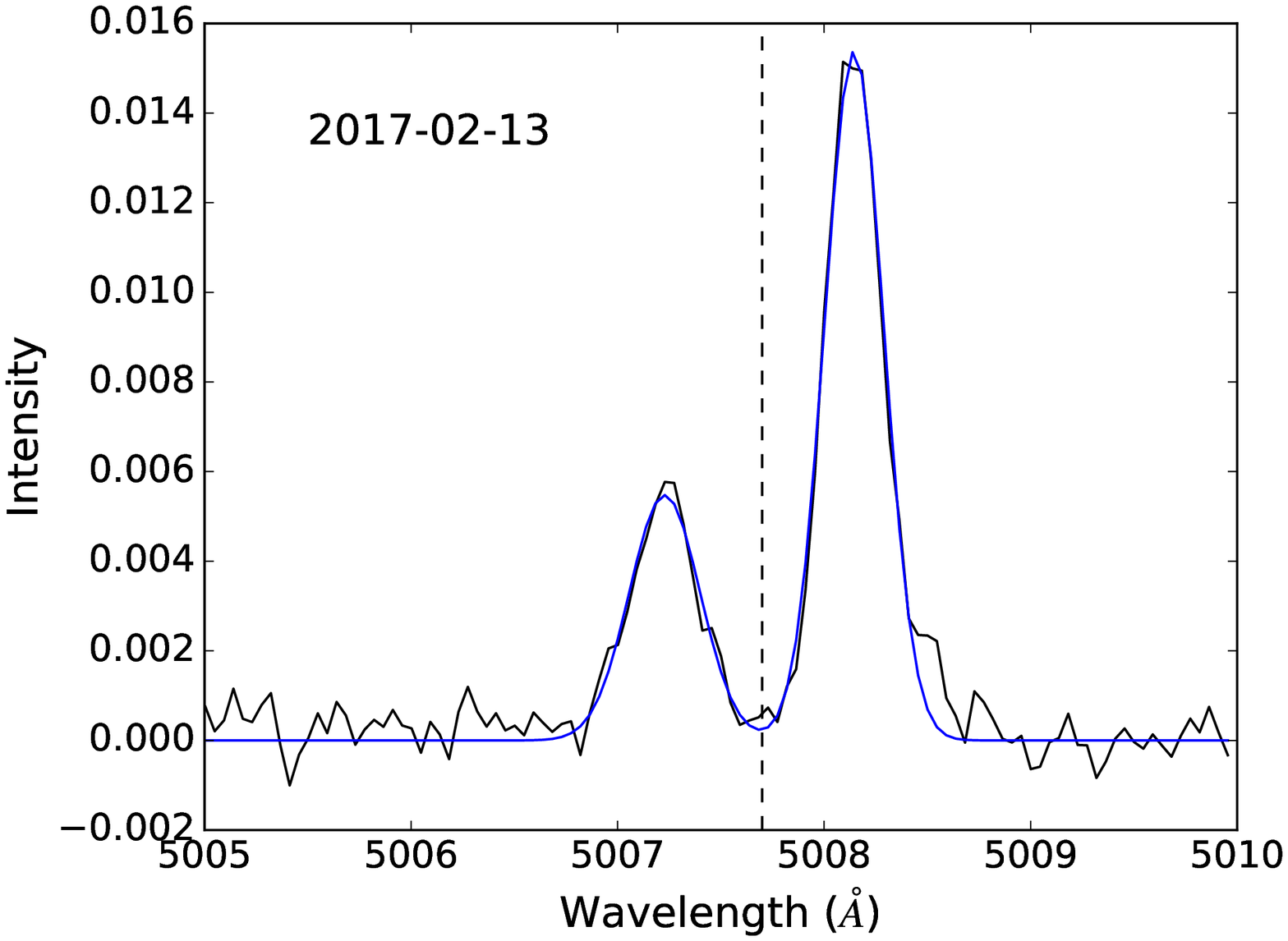}
\includegraphics[scale=0.17]{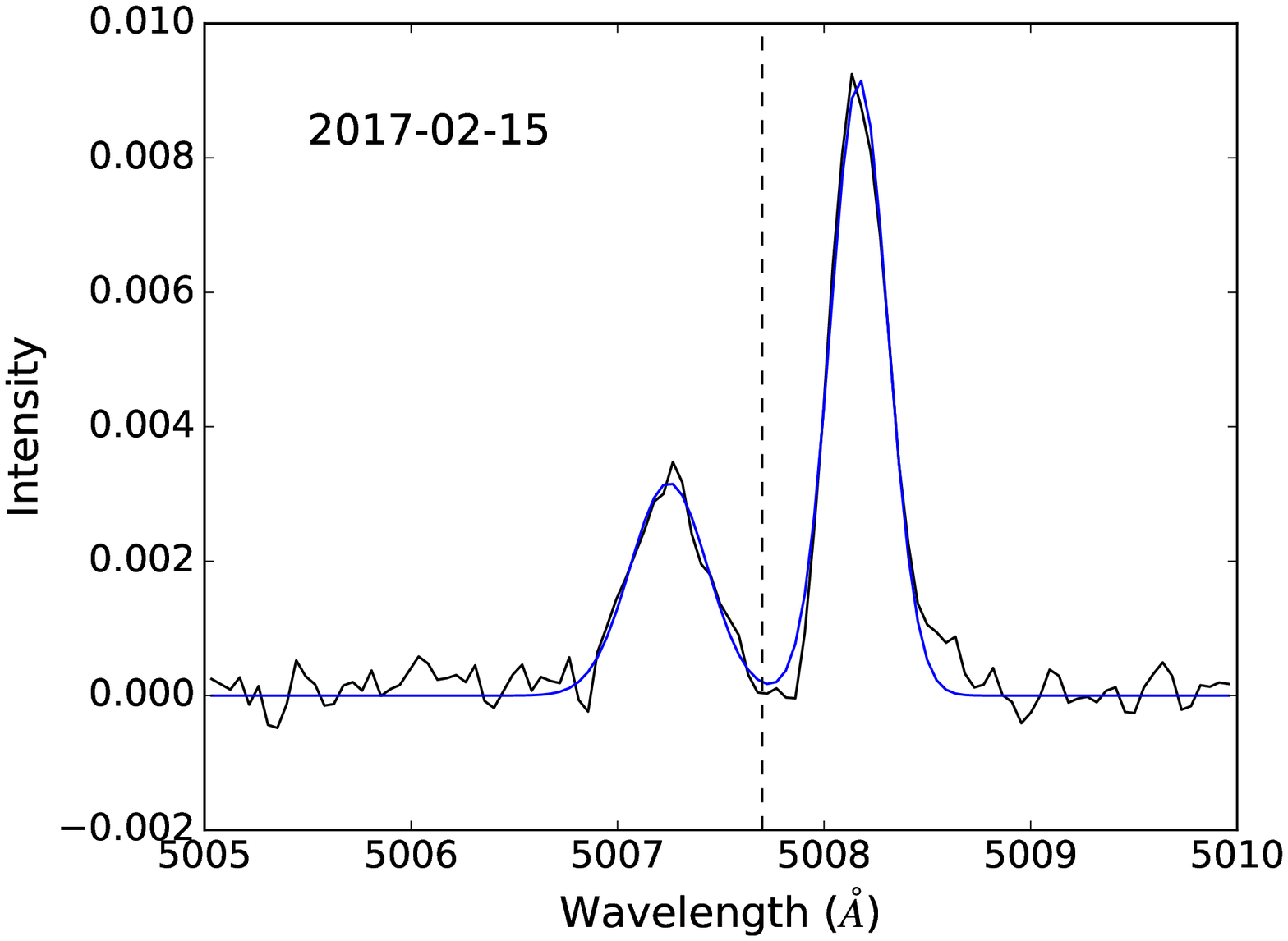}
\includegraphics[scale=0.17]{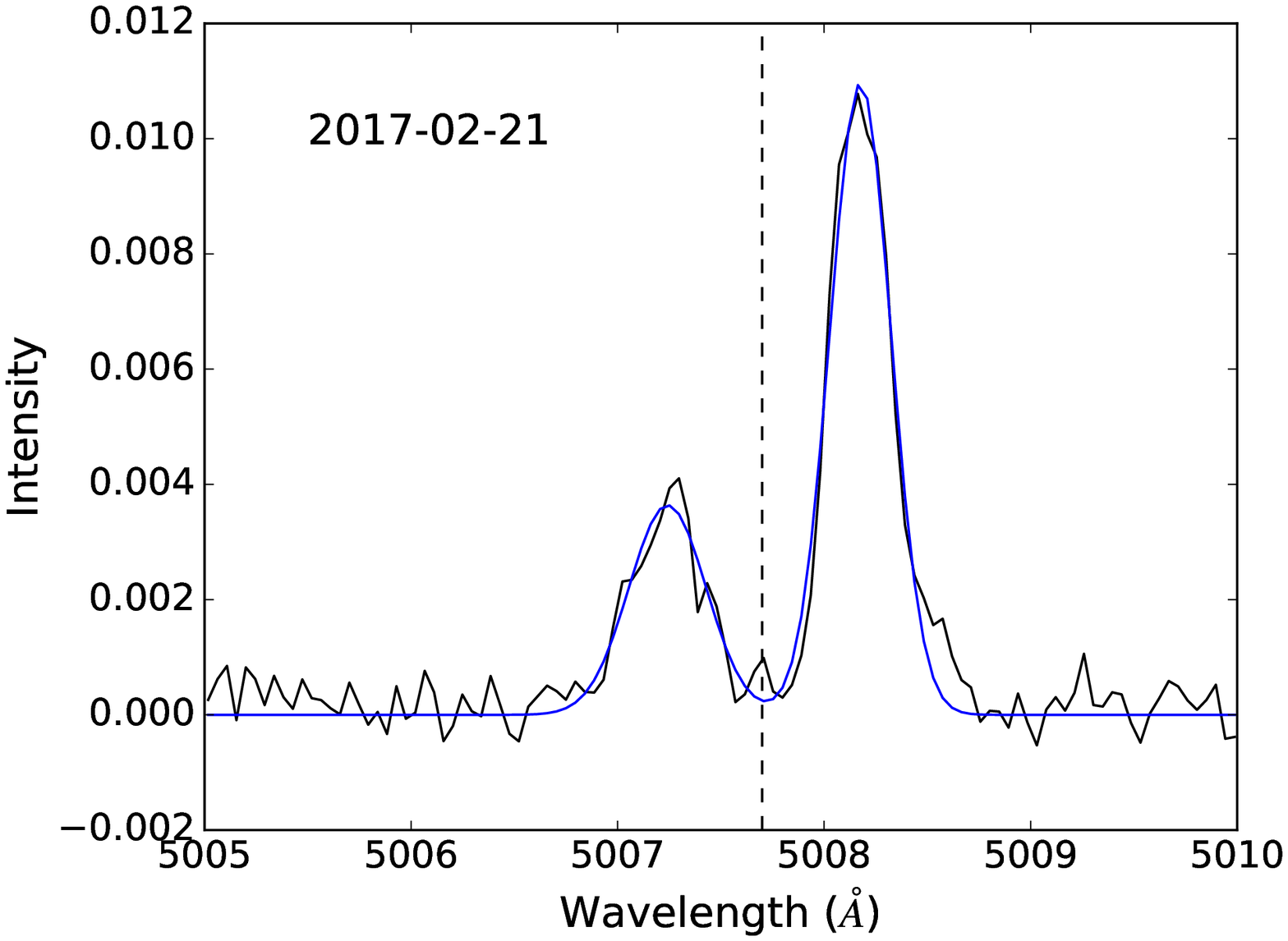}
\includegraphics[scale=0.17]{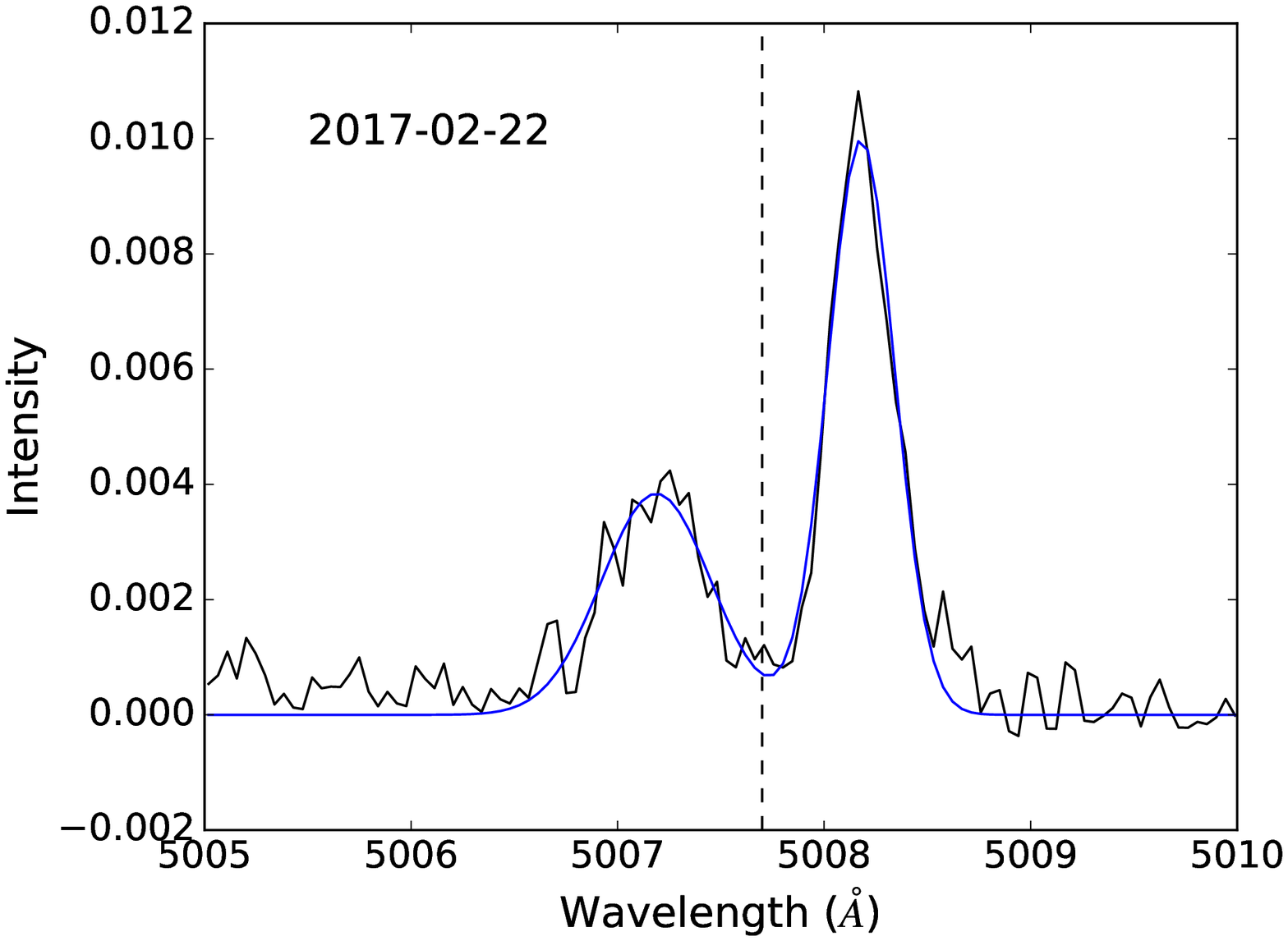}
\includegraphics[scale=0.17]{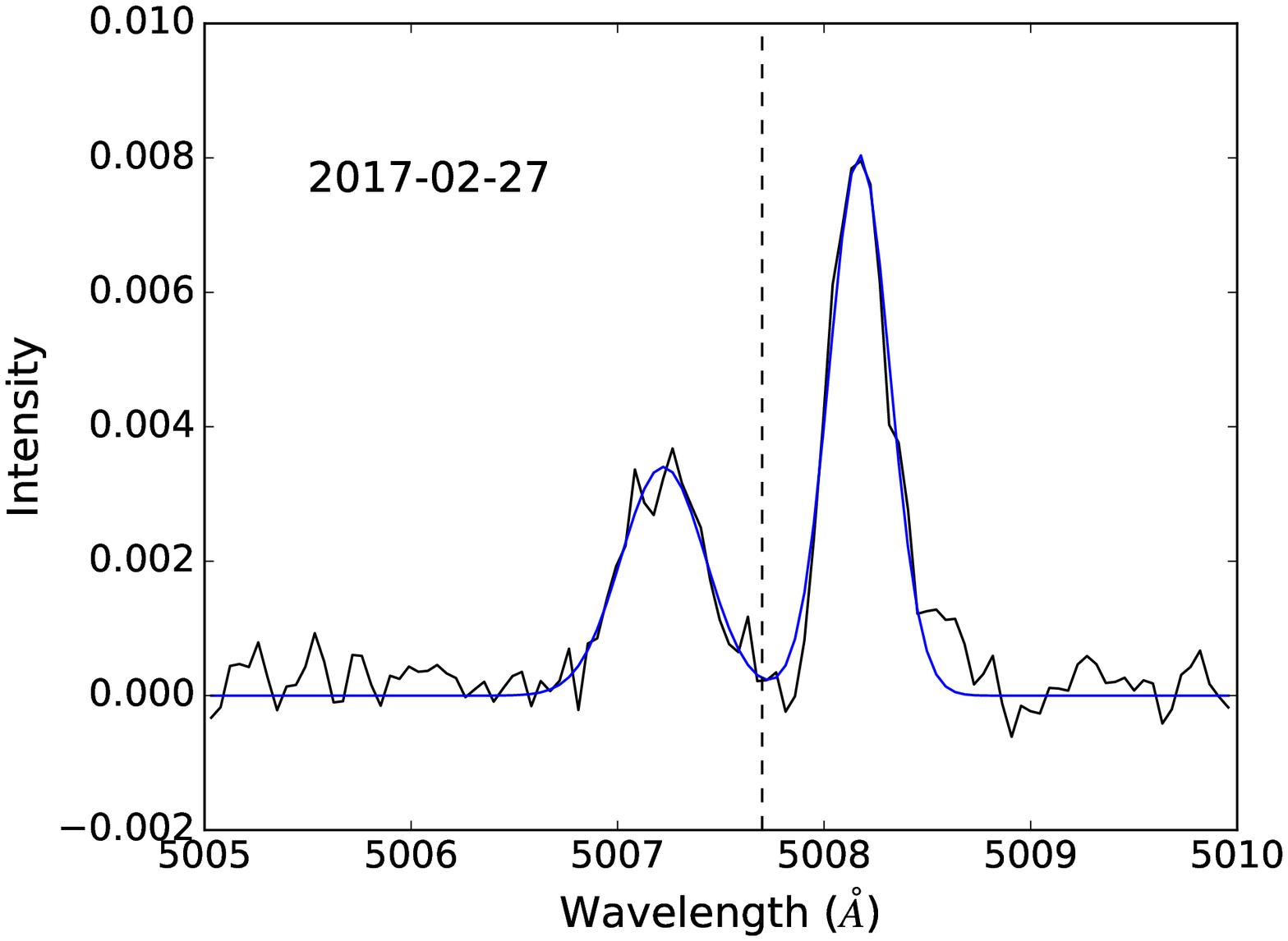}
\includegraphics[scale=0.17]{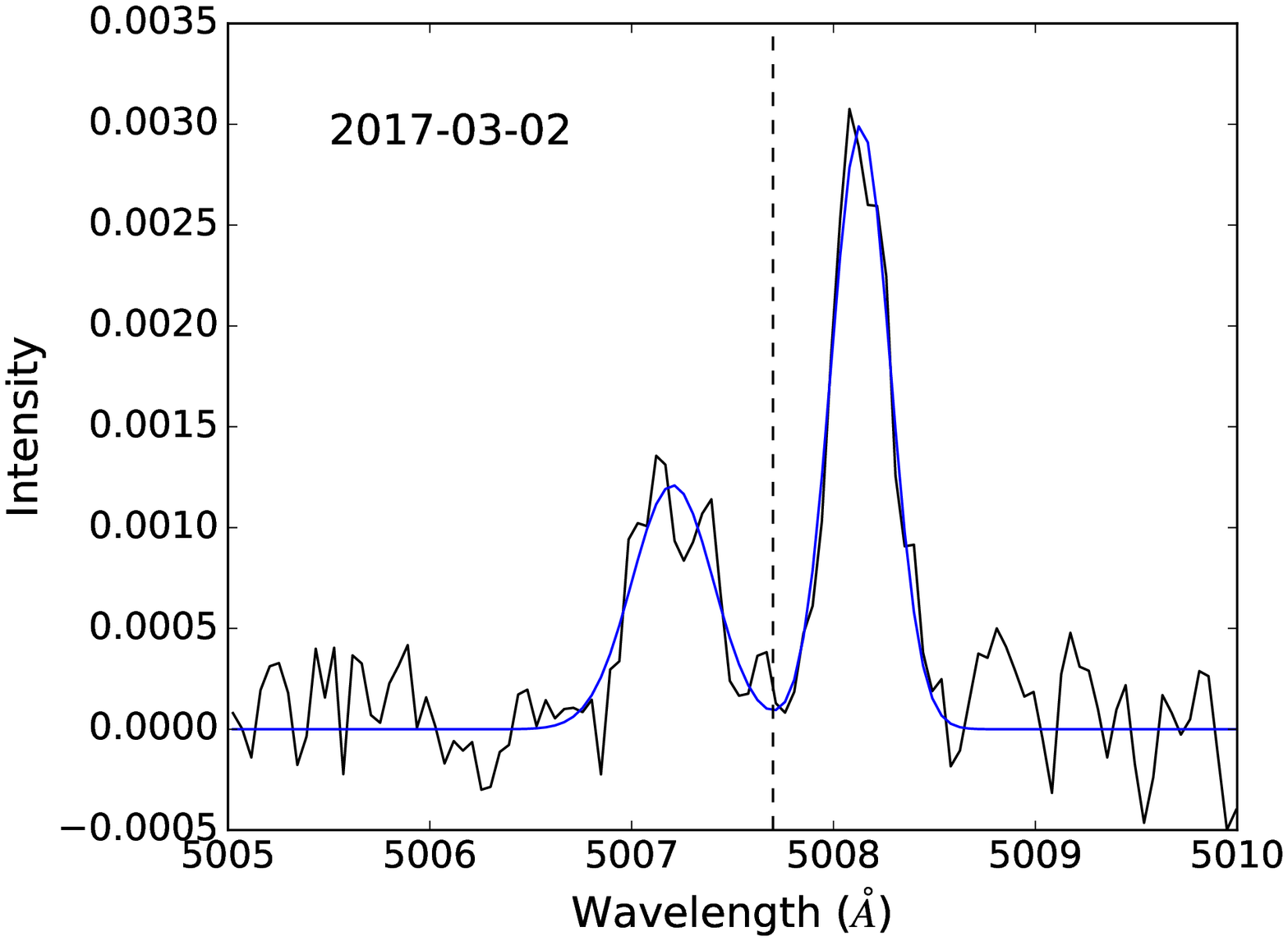}
\includegraphics[scale=0.17]{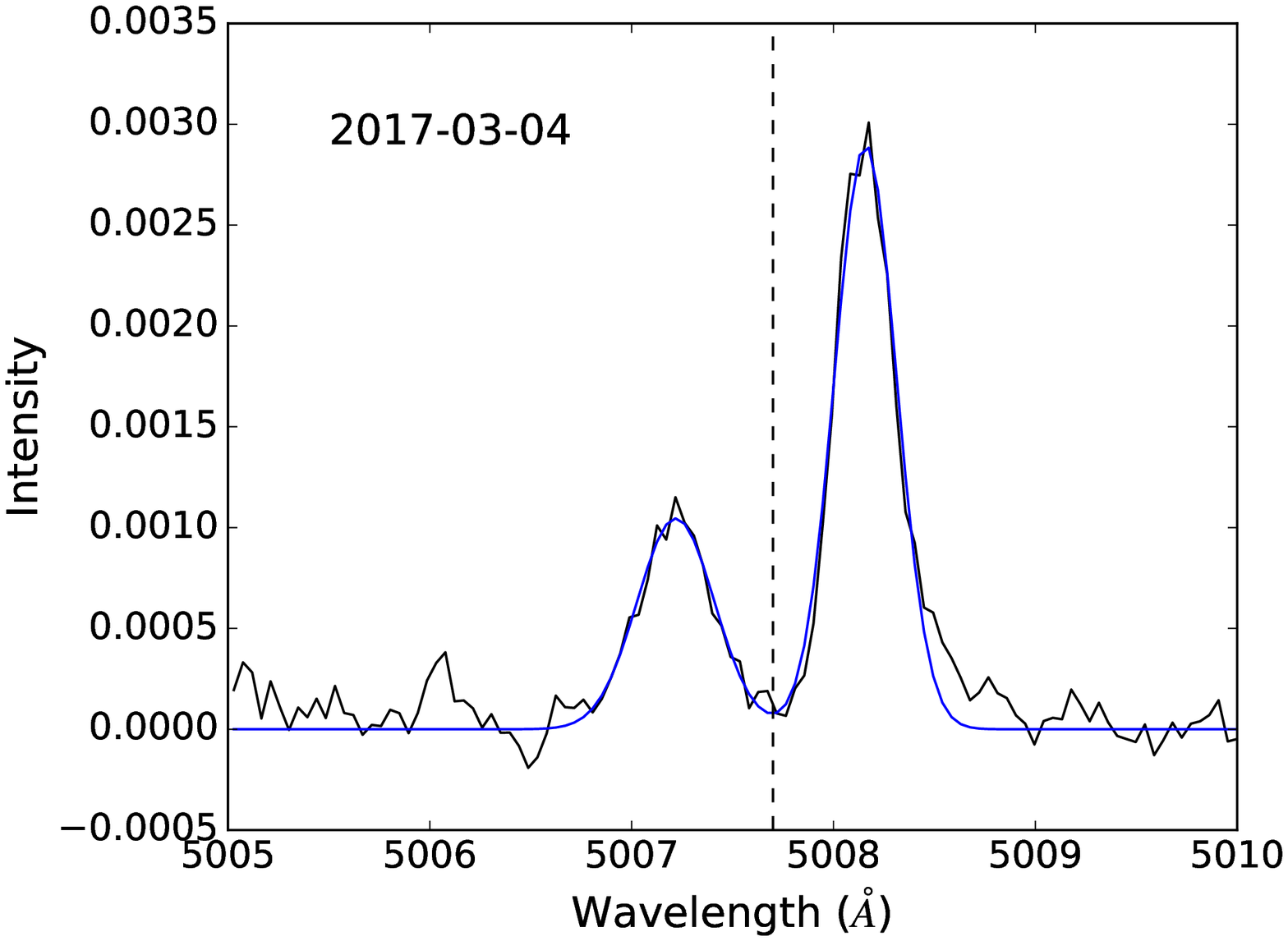}
\includegraphics[scale=0.17]{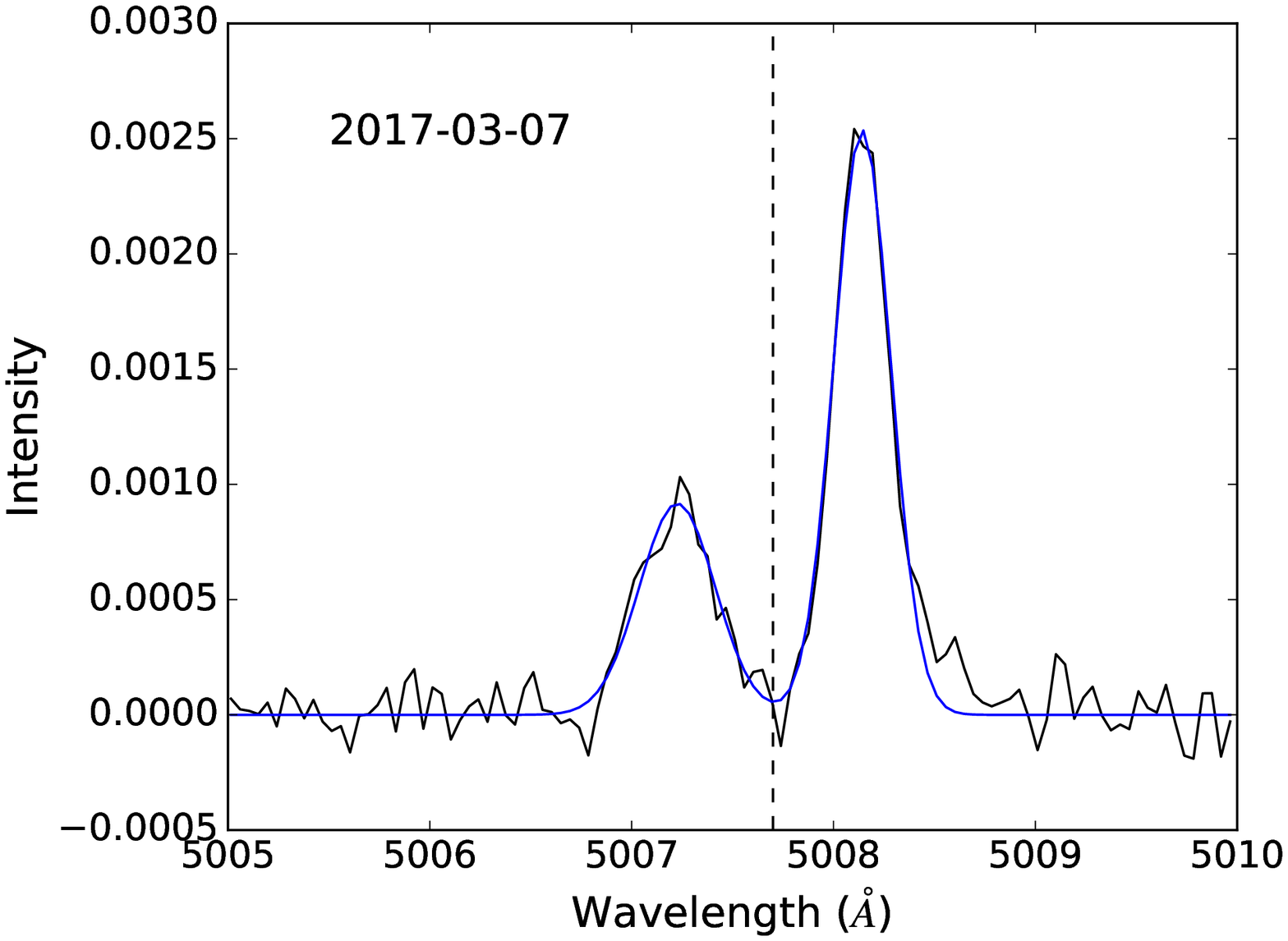}
   \end{center}
   \caption{Multiple Gaussian fits (blue) to the nebular [O~III] 5007 emission line of each spectrum (black). The dashed line corresponds to the average nebular radial velocity of 50.47 km s$^{-1}$ determined from the fits.}
   \label{fig:nebfits}
\end{figure*}

The spectrum of HIP~16566 contains several weak emission lines suitable for RV measurements that are preferable to the much broader and complex profiles of the absorption lines (see e.g. Jordan et al. 2005). We fitted Gaussians to several of the weak emission lines using the \textsc{lmfit} package (Newville et al. 2016) to determine their RV motions. Table \ref{tab:rvs} contains the RV measurements of the highest signal-to-noise emission line O~V $\lambda$4930.27\footnote{Identified from https://www.nist.gov/pml/atomic-spectra-database} that is representative of the RV measurements from all features. Figure \ref{fig:ovfits} shows the Gaussian fits to O~V $\lambda$4930 whose 1-$\sigma$ errors were used to determine the individual measurement errors in Tab. \ref{tab:rvs}. Apart from stationary nebular emission lines, all other features moved in phase with the stellar emission lines, including the absorption lines, and no features that may be from a companion could be identified. Figure \ref{fig:rvs} displays the RV time series along with the RV measurements of the velocity-resolved [O~III] $\lambda$5007 emission line. Both components of [O~III] $\lambda$5007 were fitted with Gaussians (Fig. \ref{fig:nebfits}) giving a mean nebular heliocentric RV of 50.47$\pm$0.73 km s$^{-1}$, in good agreement with 49 km s$^{-1}$ of Goldman et al. (2004). The uncertainties in the nebular measurements were determined by adding in quadrature the uncertainty from the Gaussian fit to each component. The scatter in the nebular measurements may be due to small differences in the sky position of the HRS input fibre. There is a small difference between the observed average velocity of O~V $\lambda$4930 (67.09$\pm$0.13 km s$^{-1}$) and the nebular systemic velocity. This is not unexpected for a weak wind which is present in HIP~16566 as demonstrated by the emission core of He~II $\lambda$4686 (M\'endez et al. 1988). Depending on the line formation region, a small offset of several tens of km s$^{-1}$ can occur (see e.g. M\'endez et al. 1990). 
\begin{table}
   \centering
   \caption{SALT HRS RV measurements of O~V $\lambda$4930.27.}
   \label{tab:rvs}
   \begin{tabular}{ll}
      \hline
      Julian day & RV (O~V)\\
                 & (km s$^{-1}$) \\
      \hline
   2457675.41096 & 65.79 $\pm$ 0.34\\
   2457708.33082 & 55.77 $\pm$ 0.33\\
   2457734.49451 & 61.95 $\pm$ 0.37\\
   2457766.37974 & 76.74 $\pm$ 0.38\\
   2457768.37785 & 77.14 $\pm$ 0.33\\
   2457792.31382 & 77.82 $\pm$ 0.31\\
   2457798.30426 & 74.99 $\pm$ 0.29\\
   2457800.31159 & 74.12 $\pm$ 0.48\\
   2457806.27403 & 71.87 $\pm$ 0.36\\
   2457807.28034 & 71.81 $\pm$ 0.28\\
   2457812.28064 & 68.96 $\pm$ 0.38\\
   2457815.28142 & 66.79 $\pm$ 1.06\\
   2457817.25951 & 65.46 $\pm$ 0.38\\
   2457820.25700 & 63.72 $\pm$ 0.32\\
   \hline
   \end{tabular}
\end{table}

\begin{figure*}
   \begin{center}
      \includegraphics[scale=0.84]{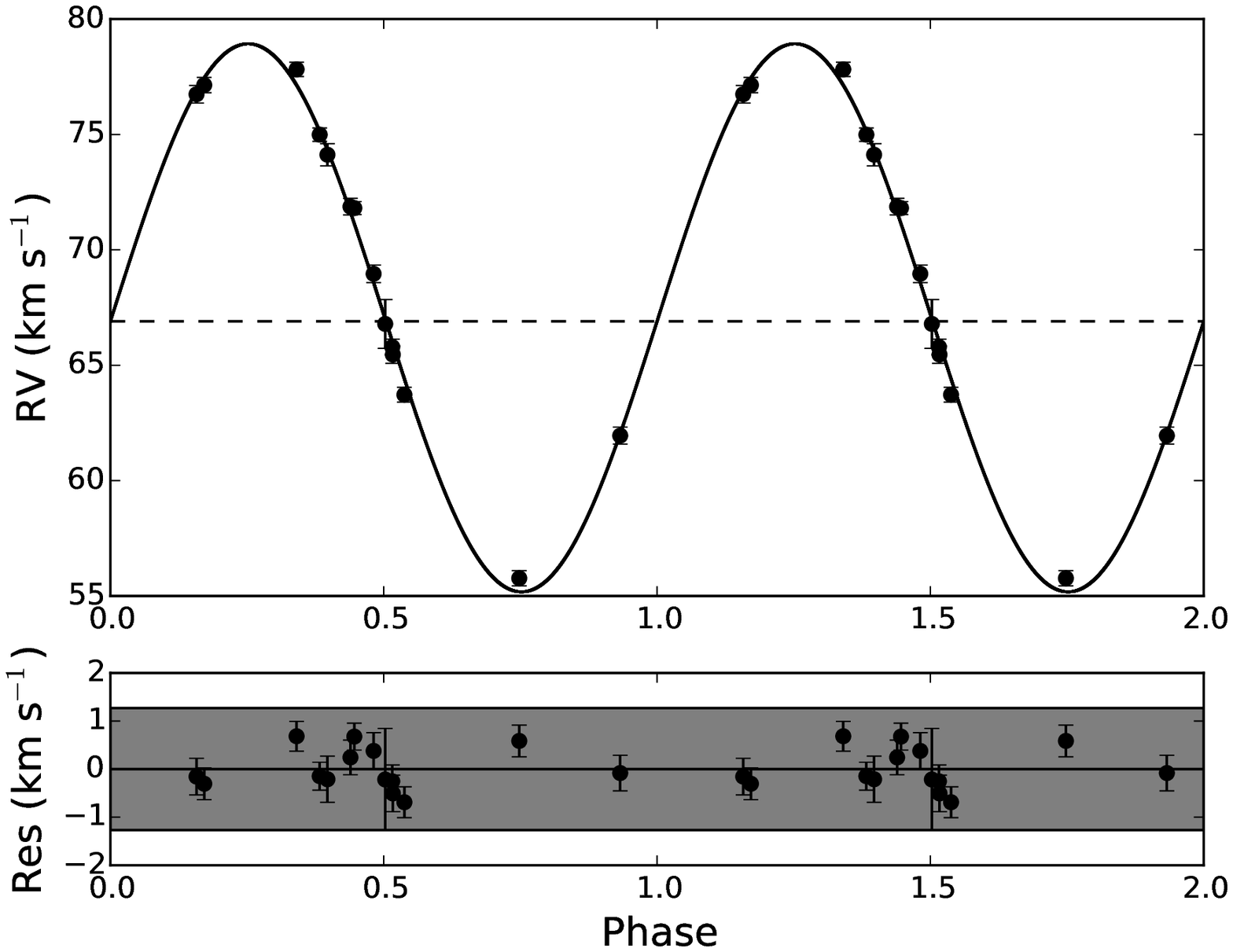}
   \end{center}
   \caption{\emph{Top panel:} SALT HRS heliocentric RV measurements of O~V $\lambda$4930 in HIP 16566 folded on the orbital period. The solid line is the Keplerian orbit fit. \emph{Bottom panel:} The residuals are within 3$\sigma$ of the fit (shaded region).}
   \label{fig:rvcurve}
\end{figure*}

   It is clear from Fig. \ref{fig:rvs} that coherent sinusoidal variability is present in the RV measurements of HIP~16566. The original RVs were pre-whitened to construct a periodogram and the most significant period was used to fit an initial Keplerian model. A least-squares minimisation method was then used to obtain the best fit. The eccentricity was fixed to zero in all our fits. Figure \ref{fig:rvcurve} displays the RV measurements folded with the orbital period of 141.6$\pm$0.8 d and Tab. \ref{tab:orbit} contains the best fit parameters. The parameter $a_1 \sin i$ denotes the distance of the primary from the centre of mass of the binary.

\begin{table}
   \centering
   \caption{Orbital parameters of HIP~16566 derived from the best-fit Keplerian orbit to O~V $\lambda$4930.}
   \label{tab:orbit}
   \begin{tabular}{lll}
      \hline
      Period (d) & 141.6 $\pm$ 0.8& \\
      $e$ (fixed) & 0.0 & \\
      $K$ (km s$^{-1}$) & 11.80 $\pm$ 0.18 & \\
      $\gamma$ (km s$^{-1}$) & 67.09 $\pm$ 0.13 & \\
      $T0$ (d) & 2457708.9 $\pm$ 0.8 & \\
      $a_1 \sin i$ (AU) & 0.1535 $\pm$ 0.0023 & \\
      $f(M)$ ($M_\odot$) & 0.0241 $\pm$ 0.0011 & \\
      RMS residuals (km s$^{-1}$) & 0.42 & \\
      \hline
   \end{tabular}
\end{table}

The errors in the orbital parameters were computed using a Monte Carlo method where the data were randomised by adding Gaussian noise to the RVs with the 1-$\sigma$ value taken from the individual errors. We created 10000 randomised Keplerian orbits fit to the data and the errors were then calculated by taking the standard deviation of the parameters of all Keplerian models. The uncertainty in the orbital period and date of minimum RV ($T0$) are formally 0.8 d, but could be as large as 1.5 d given the incomplete phase coverage. Similarly, while a circular orbit provides an excellent fit to the data, the current phase coverage does not allow us to formally exclude mildly eccentric orbits up to $e\sim$0.1. The best fit Keplerian orbit results in root-mean-square (RMS) residuals of 0.42 km s$^{-1}$. 

\subsection{Nature of the companion}
\label{sec:comp}
Adopting a mass of $M_1=0.555\pm0.030$ $M_\odot$ for the primary (Sect. \ref{sec:ngc1360}) we tabulate the possible companion masses in Tab. \ref{tab:masses}. While there are difficulties in adopting masses for CSPNe from single-star evolutionary tracks (see section 4.2.3 of Moe \& De Marco 2006 and Moe \& De Marco 2012), we prefer this approach over assuming the standard 0.6 $M_\odot$. If we further assume that the orbital plane of the binary and the nebula are coplanar, as has been demonstrated for several post-CE PNe (Hillwig et al. 2016 and ref. therein), the likely companion mass would be $M_2=0.68^{+0.62}_{-0.21}$ $M_\odot$ for $i=30\pm10$ deg derived from the nebula (Goldman et al. 2004). The orbital separation between the two stars is given by $a=a_1+a_2$ and $a_1/a_2=M_2/M_1$, where $a_i$ denotes the separation of each star from the centre of mass of the binary, $M_1=0.555$ and $M_2=0.679$ $M_\odot$. Under these assumptions $a=0.307+0.251=0.558$ AU corresponding to 120 $R_\odot$.

There are two possible companion types that could satisfy the permissible companion mass range: a cool main-sequence companion, or a white dwarf of similar or greater mass than the primary.

\begin{table}
   \caption{Companion masses $M_2$ for $M_1=0.555\pm0.03$ $M_\odot$ and a variety of orbital inclination angles. }
   \label{tab:masses}
   \centering
   \begin{tabular}{cl}
   \hline
   Inclination ($^\circ$) & $M_2$ ($M_\odot$) \\
   \hline
   10 & $5.760\pm0.530$ \\
   20 & $1.301\pm0.108$ \\
   30 & $0.679\pm0.046$ \\
   40 & $0.465\pm0.028$ \\
   50 & $0.364\pm0.028$ \\
   60 & $0.308\pm0.024$ \\
   70 & $0.278\pm0.021$ \\
   80 & $0.261\pm0.020$ \\
   90 & $0.256\pm0.015$ \\
   \hline
   \end{tabular}
\end{table}

\begin{figure}
   \begin{center}
      \includegraphics[scale=0.4]{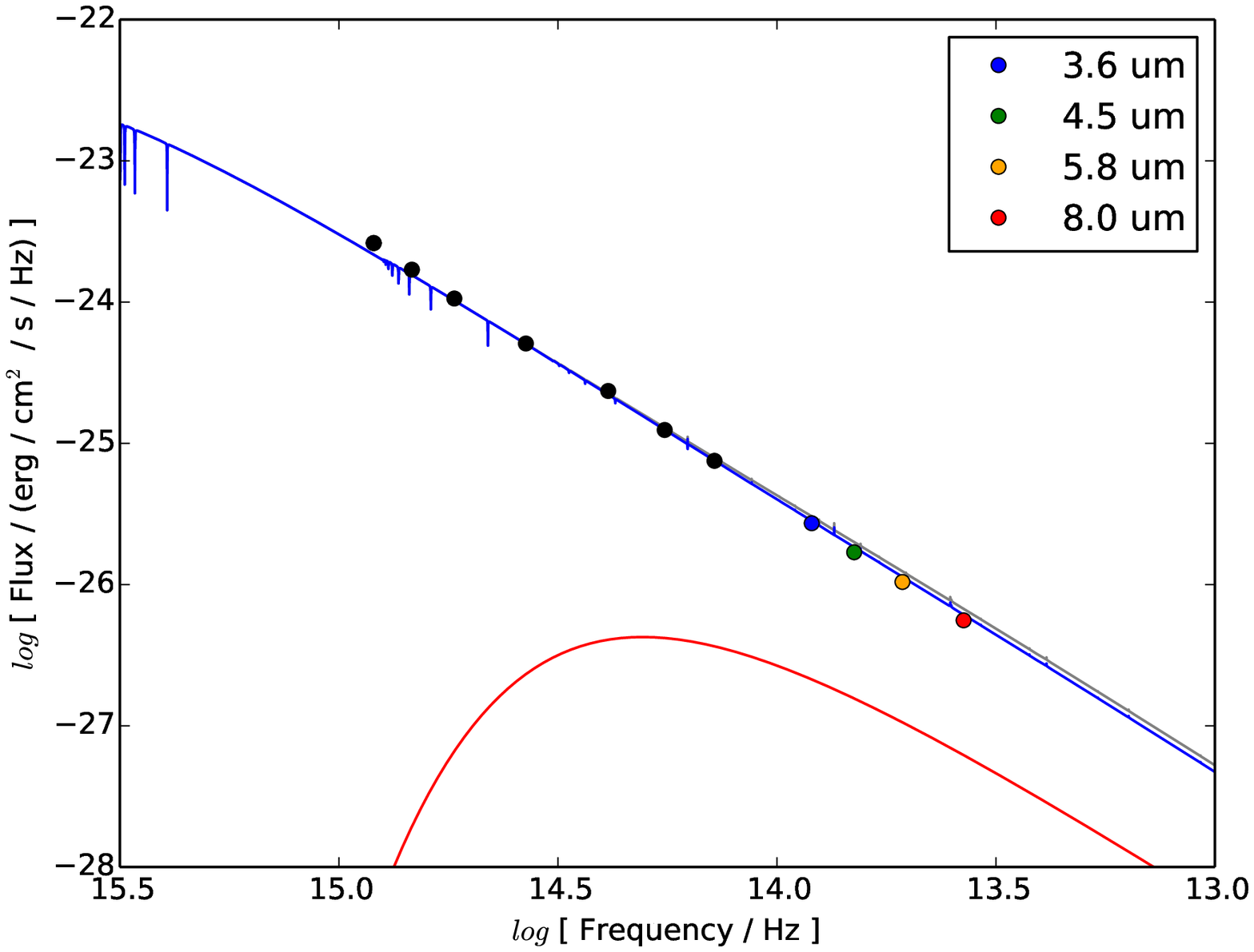}\\
      \includegraphics[scale=0.4]{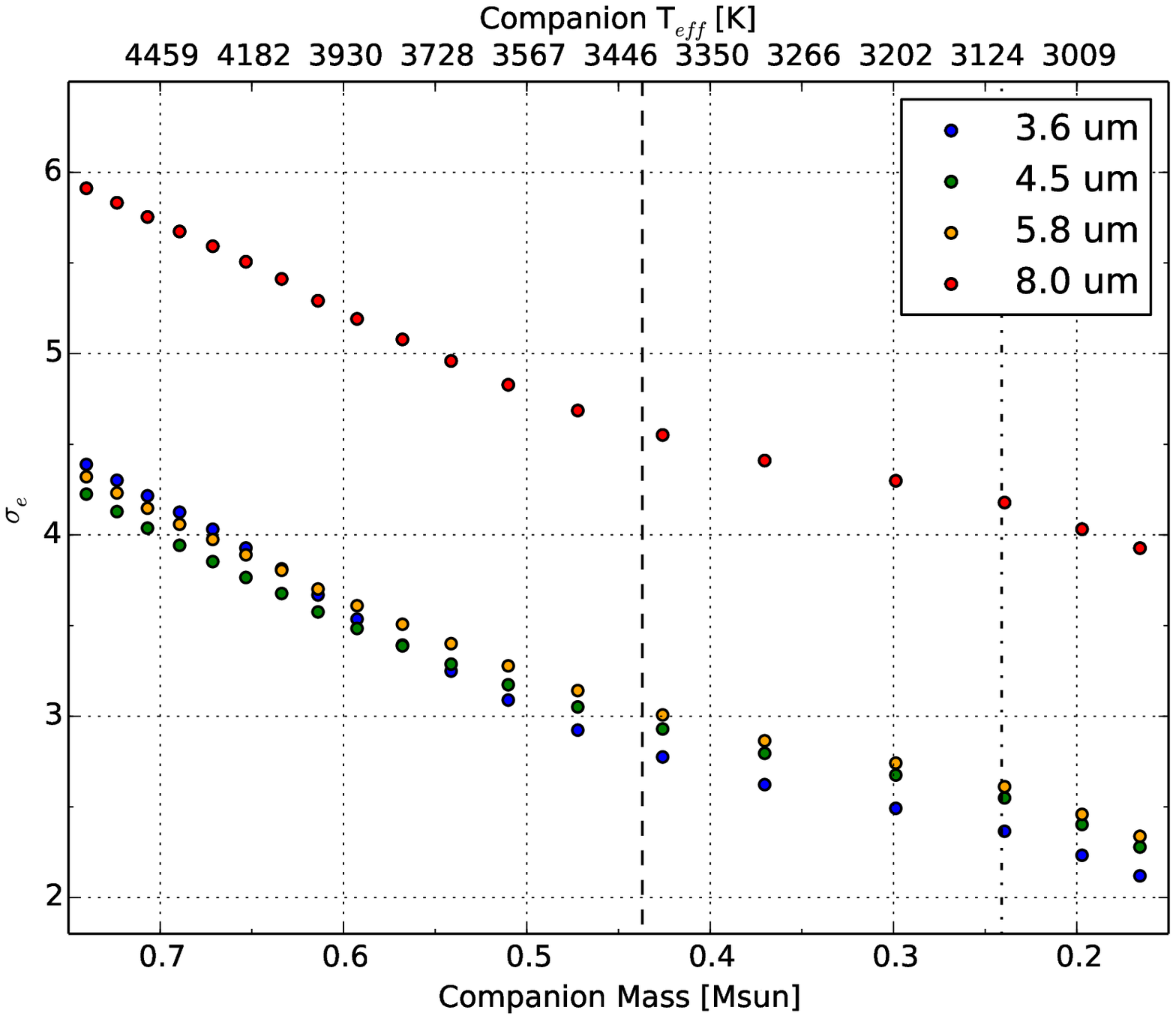}
   \end{center}
   \caption{\emph{Top panel:} Observed photometry of HIP~16566 from Bil{\'{\i}}kov{\'a} et al. 2012, a model SED of the primary (blue line), a hypothetical 0.45 $M_\odot$ main-sequence companion blackbody (red line, $T_\mathrm{eff}=3446$ K) constructed using software developed by Astropy Collaboration et al. (2013) (red line) and the combined SED (grey line). \emph{Bottom panel:} Mid-infrared excess $\sigma_e$ of the SED combining the primary SED and hypothetical main-sequence companions over the observed \emph{Spitzer} IRAC band fluxes measured by Bil{\'{\i}}kov{\'a} et al. (2012). The vertical dashed line corresponds to $M_2$=0.437, derived from the upper limit on the inclination of 40 deg (Goldman et al. 2004), whereas the dash-dotted line corresponds to $M_2=0.241$, the lowest companion mass permissible by the mass function.}
   \label{fig:bb}
\end{figure}

To determine some quantitative limits on the nature of the companion, we have reassessed the photometry used by Bil{\'{\i}}kov{\'a} et al. (2012) to search for a cool companion. Flux zeropoints for the optical and NIR magnitudes were taken from Bessell (1979) and Cohen et al. (2003). For the primary spectral energy distribution (SED) we used a synthetic spectrum of a pure hydrogen star with log $g=5.7$ cm s$^{-2}$ and $T_\mathrm{eff}=105$ kK from Rauch \& Deetjen (2003). The primary radius was assumed to be $R=0.135$ $R_\odot$, recalculated from Herald \& Bianchi (2011) at $d=421$ pc. We scaled the primary SED first by $S_p=R^2/d^2$, and then by another factor $S_m$ to match the observed $VIJK_s$ photometry using a least-squares method. For the potential main-sequence companion SED we used magnitudes from the Phoenix BT-Settl models (Baraffe et al. 2015) and a $T_\mathrm{eff}$-radius-mass relation for MS stars with $Z=0.014$ (Bressan et al. 2012; Chen et al. 2014). Taking the combined primary and secondary SEDs for different secondary masses, we calculated the excess $\sigma_e$ in the combined flux compared to the \emph{Spitzer} IRAC flux in the four bands (Bil{\'{\i}}kov{\'a} et al. 2012). We define $\sigma_e$ as the combined primary and main sequence star flux minus the IRAC fluxes, divided by the 1-$\sigma$ errors of each IRAC flux. The main-sequence SEDs were scaled to the distance allowed by the derived error margin ($421_{-131}^{+123}$ pc) that gave the smallest value of $\sigma_e$. Figure \ref{fig:bb} shows an example of the SEDs and the observed photometry, as well as the excess $\sigma_e$ as a function of companion mass. If we restrict the possible range of orbital inclinations to $20<i<40$ deg (Goldman et al. 2004), which also contains the 33.7$^{+11.7}_{-23.7}$ deg inclination of the candidate ring of LIS (Sect. \ref{sec:lis}), then the mass function combined with Fig. \ref{fig:bb} means that we can exclude a main-sequence companion at the 4.5$\sigma$ level. Furthermore, we can exclude at the 4$\sigma$ level all permissible main-sequence companions ($M_2\ge0.241$ $M_\odot$).

The chance of HIP~16566 hosting a cool main-sequence companion is therefore remote. This is supported by the lack of harder X-ray emission that would originate from the corona of a cool companion (Sect. \ref{sec:ngc1360}). The only remaining possible companion is a WD of similar or greater mass, though observing such companions is very difficult. In some cases both WDs may be observable due to their similar luminosities (e.g. Santander-Garcia et al. 2015), but we find no evidence for a WD of similar luminosity in the HRS spectra. These facts combined with the companion mass range of $M_2=0.68^{+0.62}_{-0.21}$ suggest the most likely scenario for HIP~16566 is a more massive WD companion as in other post-CE PNe Fleming~1 (Boffin et al. 2012) and NGC~5189 (Manick et al. 2015). In these cases the bright primary dominates the spectral energy distribution, making it very difficult to detect any trace of a companion (even at UV wavelengths). The large range of companion masses allowed by the mass function comes from the relatively large 10 deg error in the nebula inclination (Goldman et al. 2004), precluding us from giving any accurate physical parameters for the secondary.

\section{Post-common-envelope status of NGC~1360}
\label{sec:postCE}
\begin{figure}
   \begin{center}
      \includegraphics[scale=0.5]{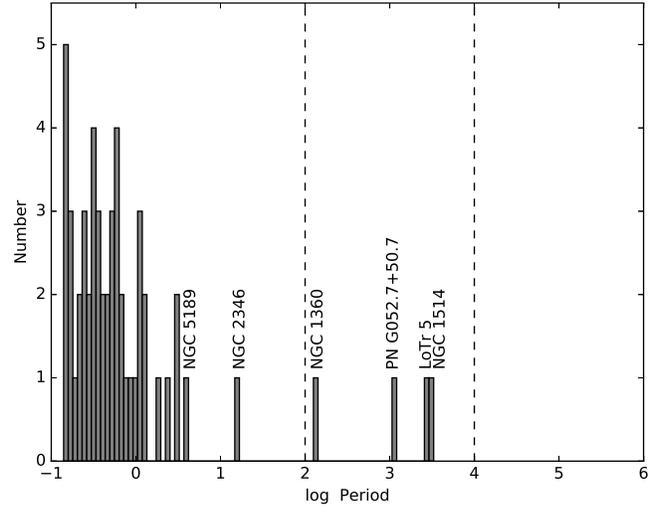}
   \end{center}
   \caption{Orbital period distribution of binary central stars of PNe. Longer period systems are labelled and the dashed lines indicate the range populated by post-AGB binaries (Van Winckel et al. 2009; Gorlova et al. 2014; Manick et al. 2017).}
   \label{fig:pdist}
\end{figure}

The 142 d orbital period of the double-degenerate (DD) binary nucleus of NGC~1360 is remarkable amongst PNe. Figure \ref{fig:pdist} shows the current orbital period distribution of PNe.\footnote{The close binaries in the orbital period distribution include those listed in appendix A of Manick et al. (2015) with the addition of J193110888$+$4324577 (De Marco et al. 2015), Hen~2-155 (Jones et al. 2015), Th~3-15 (Soszy{\'n}ski et al. 2015; Hlabathe 2015), H~2-22, PPA~1741-2538, H~2-13 and PHR~1805-2520 (Hlabathe 2015), and HaTr~7 (Hillwig et al. 2017).} The binary central star of NGC~1360 is the only one located between the orbital periods of NGC~2346 (16 d , M\'endez \& Niemel{\"a} 1981) and PN G052.7$+$50.7 (1105 d, Van Winckel et al. 2014). 

A strong observational bias towards the shorter periods is certainly affecting the orbital distribution, but HIP16566 belongs to those with longer orbital periods. The observational bias is much less of a problem in the evolutionary phase preceeding the PN phase, namely among post-AGB stars. The post-AGB binaries fall between post-CE systems that were subject to a larger degree of inspiralling on one hand, and binaries with log $P>>3.0$ that only encountered wind interaction on the other hand. The orbital period of NGC~1360 clearly lies amongst the commonly detected post-AGB binaries (Van Winckel et al. 2009; Gorlova et al. 2014; Manick et al. 2017). Clearly HIP~16566 is not a canonical post-AGB binary (Van Winckel 2003), but similarly short orbital periods to HIP~16566 are found amongst the post-AGB binaries BD$+$46$^\circ$442 ($P=140$ d, Gorlova et al. 2012), SAO~173329 ($P=116$ d, Van Winckel et al. 2000) and IRAS19157-0247 ($P=119$ d, Van Winckel et al. 2009). 

Its location at the lower end of the post-AGB binary orbital period range suggests that HIP~16566 experienced one or more common-envelope interactions. A dramatic reduction in the orbit is clearly demonstrated by the contrast between the AGB star radii of 274 and 523 $R_\odot$ for the component WD masses of 0.55 and 0.68 $M_\odot$ (Eq. (5) of Rappaport et al. 1995), respectively, and the current separation between the two stars of 120 $R_\odot$ (assuming $i=30$ deg, see Sect. \ref{sec:comp}). Population synthesis models also predict post-CE orbital periods near log $P=2.15$ for single degenerate systems (figure 16 of Nie et al. 2012; Moe \& De Marco 2012). In the case of DD systems the Han (1998) models predict only marginally non-zero numbers of CO+CO WDs for $P>10$ d. Further comparisons with PSM are premature at this stage, especially considering the lack of similarly long period DD systems to test and refine model predictions. There may be other possible evolutionary scenarios that could explain NGC~1360 such as grazing envelope evolution (Soker 2015). Apart from the 30.09 d DA+DB binary PG 1115$+$166 (Maxted et al. 2002), the search for DD binaries has been heavily biased towards very short period systems (e.g. Napiwotzki et al. 2001). The long orbital period and DD nature of HIP~16566 makes it an important test case with which to improve PSM.

Apart from the predictions made by PSM, there are potentially three main pieces of corroborating evidence that support a post-CE status for NGC~1360:
\begin{enumerate}
   \item Low-ionisation structures feature prominently along the polar axis (Goldman et al. 2004; Garc{\'{\i}}a-D{\'{\i}}az et al. 2008) and in the newly identified ring (Sect. \ref{sec:lis}). It is the ring configuration, in particular, that is notable amongst other post-CE PNe, namely Sab~41 (Miszalski et al. 2009b), NGC~6337 (Garc{\'{\i}}a-D{\'{\i}}az et al. 2009), The Necklace (Corradi et al. 2011) and Fleming~1 (Boffin et al. 2012). Furthermore, many other post-CE show nebulae dominated by LIS (Miszalski et al. 2011c; Jones et al. 2014; Manick et al. 2015), or alternatively in isolated or point-symmetric knots that could be interpreted as rings in a more evolved state that have been heavily disrupted (Miszalski et al. 2009b).
   
   \item NGC~1360 falls on the lower, optically thin trend of the Frew (2008) surface-brightness radius relation statistical distance indicator which is populated by several post-CE PNe with low ionised masses (see also Frew et al. 2016). Possible explanations for the low ionised mass could include a reduction in nebula mass either before the CE was ejected via jets (Tocknell et al. 2014) or after the CE was ejected via a circumbinary disk (Kashi \& Soker 2011; Kuruwita et al. 2016).

   \item The low density patches discussed by Goldman et al. (2004), see also Sect. \ref{sec:ngc1360} and Fig. \ref{fig:lis}, are qualitatively similar to the same seen in the post-CE PN A~65 (Huckvale et al. 2013). Simulations of post-CE PNe indicate that gas can fall back towards the central star (e.g. Kashi \& Soker 2011) and this could potentially create the appearance of such low density patches as falling gas reclaims previously cleared out regions. Low density patches are unusual amongst PNe and we encourage further modelling to understand their formation.

\end{enumerate}
      
\section{Conclusions}
\label{sec:conclusion}

The central star of NGC~1360, HIP~16566, has been the beguiling subject of numerous observational efforts that unsuccessfully tried to detect signatures of a magnetic field (Jordan et al. 2005; Leone et al. 2011, 2014; Jordan et al. 2012) and a binary companion (M{\'e}ndez \& Niemel{\"a} 1977; Wehmeyer \& Kohoutek 1979; Af{\v s}ar \& Bond 2005; Bil{\'{\i}}kov{\'a} et al. 2012). As part of a large systematic search for long period CSPNe with SALT HRS, we have used SALT HRS spectra with resolving power $R\sim43000$ to definitively prove the presence of a binary companion to HIP~16566. Our main conclusions are as follows:

\begin{enumerate}
   \item The SALT HRS RV time series measured from the O~V $\lambda$ 4930\AA\ emission line in HIP~16566 was fit with a circular Keplerian orbit with an orbital period of 142 d and an amplitude of 11.8 km s$^{-1}$. Key to this discovery was the excellent stability of HRS demonstrated by the low RMS of the residuals from the Keplerian fit of 0.42 km s$^{-1}$, as well as the $\sigma=$0.73 km s$^{-1}$ scatter in nebular [O~III] RV measurements.
   \item Assuming the binary is coplanar with the nebula (e.g. Hillwig et al. 2016), the orbital inclination is well constrained to be $i=30\pm10$ deg from the Goldman et al. (2004) spatio-kinematic study of the prolate ellipsoidal nebula. Further supporting this is the 33.7$^{+11.7}_{-23.7}$ deg inclination determined from fitting an ellipse to a newly discovered ring of 120 candidate LIS identified from archival VLT FORS2 imaging. 
   \item No features from the companion were visible in the SALT HRS spectra. We investigated whether the companion could be a cool main-sequence star using the photometry from Bil{\'{\i}}kov{\'a} et al. (2012). Analysis of the flux contributions that main-sequence companions would make to a synthetic spectrum of a primary at \emph{Spitzer} IRAC wavelengths show that we can rule out a main-sequence companion at the 4.5$\sigma$ level for $20<i<40$ deg and at the 4$\sigma$ level all permissible main-sequence companions ($M_2\ge0.241$ $M_\odot$).
   \item The companion is most likely a WD of similar or greater mass ($M_2=0.68^{+0.62}_{-0.21}$ $M_\odot$). We find no evidence of a WD with similar luminosity to HIP~16566 in the HRS spectra. The companion is therefore likely to be a more massive evolved WD like those present in the post-CE PNe Fleming~1 (Boffin et al. 2012) and NGC~5189 (Manick et al. 2015). Observing such a companion would be very difficult, even with future X-ray and UV observations, given its expected very low luminosity compared to HIP~16566. 
   \item NGC~1360 appears to have gone through one or more CE interactions with the orbital period (log $P=2.15$) at the long end of the predictions for single degenerate post-CE binaries (e.g. Nie et al. 2012). While only marginally non-zero numbers of long period CO+CO WD binaries are predicted for $P>10$ d (Han 1998), the relatively rapid discovery of the binary nucleus of NGC~1360 since starting this SALT HRS survey, especially in a very well-studied nearby PN, may indicate they may be more common. If verified with similar future discoveries, this could have important implications for the overall numbers of DD binaries predicted by PSM, and perhaps even result in an increase in the predicted number of type Ia supernova progenitors. In any case, a fundamental goal of our ongoing SALT HRS survey for long period binary CSPNe is to directly constrain such probabilities of PSM observationally.
   \item Several nebular features also support a post-CE origin for NGC~1360: (a) The new ring of candidate LIS we discovered in Sect. \ref{sec:lis} is typical of other post-CE PNe (e.g. The Necklace, Corradi et al. 2011), (b) Its position on the optically thin trend of the Frew (2008) statistical distance indicator where other post-CE are also located, and (c) Low-density patches as in the post-CE PN A~65 (Huckvale et al. 2013) that could be due to fallback of material after CE ejection (e.g. Kashi \& Soker 2011).
   \item The source of X-rays detected by \emph{Chandra} in NGC~1360 (Kastner et al. 2012; Montez et al. 2015) remains uncertain. The soft 0.27 keV emission, too soft to originate from a corona of a main-sequence companion (Montez et al. 2010), is consistent with the lack of photometric evidence for a main-sequence companion (Sect. \ref{sec:comp}). The most likely origin would be photospheric emission from HIP~16566 or the evolved WD companion (e.g. Guerrero et al. 2011). The origin of the X-ray emission from CSPNe in the rest of the \emph{Chandra} sample (Montez et al. 2015) remains more enigmatic. We have started monitoring some of those accessible to SALT and hope to provide binary-related constraints in the future.
   \item The close match between the 11.8 km s$^{-1}$ amplitude of the HIP~16566 RV curve and the 12 km s$^{-1}$ scatter determined by Af{\v s}ar \& Bond (2005) reinforces the mandatory role of high resolution spectra necessary to find such low-amplitude binaries (see also Van Winckel et al. 2014; Jones et al. 2017). With high-resolution observations we expect many of the objects found to be variable by De Marco et al. (2004, 2007), Sorensen \& Pollacco (2004) and Af{\v s}ar \& Bond (2005) could indeed turn out to be long-period binaries. These long period binaries may be greater in number than PSM currently predict and now, more than ever before, a binary fraction of at least 40--50\% in PNe (e.g. De Marco et al. 2009) seems within reach.
\end{enumerate}

\section{Acknowledgements}
The spectroscopic observations reported in this paper were obtained with the Southern African Large Telescope (SALT). We are grateful to our SALT colleagues for maintaining the telescope facilities and conducting the observations. We thank A. Kniazev for making available his HRS pipeline data products and for assistance in reprocessing some observations. BM thanks the Institute of Astronomy at KU Leuven and the Nicolaus Copernicus Astronomical Center (NCAC) for their hospitality during the final stages of this work. BM thanks S. Mohamed for valuable discussions, C. Ga{\l}an for local support during his NCAC visit, H. Todt for assistance with line identifications and M. M. Miller Bertolami for assistance with his models. This study has been supported in part by the Polish MNiSW grant 0136/DIA/2014/43, and NCN grant DEC-2013/10/M/ST9/0008. Polish participation in SALT is funded by grant No. MNiSW DIR/WK/2016/07. HVW, DK and RM acknowledge the support of the KU Leuven contract GOA/13/012. DK acknowledges the support by the Fund for Scientific Research of Flanders grant G.OB86.13. RM acknowledges support of the BR/143/A2/STARLAB fund by the Belgian Science Policy Office. This research made use of Astropy, a community-developed core Python package for Astronomy (Astropy Collaboration et al. 2013), and SAOImage \textsc{ds9}, developed by Smithsonian Astrophysical Observatory. IRAF is distributed by the National Optical Astronomy Observatory, which is operated by the Association of Universities for Research in Astronomy (AURA) under a cooperative agreement with the National Science Foundation.

\appendix
\newpage
\section{Candidate LIS coordinates}
\label{sec:appendix}
Table \ref{tab:coords} lists the coordinates of the 120 LIS candidates identified in Sect. \ref{sec:lis}.
\begin{table}
   \flushleft
   \caption{Coordinates of the LIS candidates.}
   \label{tab:coords}
   \begin{tabular}{llll}
      \hline
      RA ($^\circ$) & Dec. ($^\circ$) & RA ($^\circ$) & Dec. ($^\circ$) \\
      \hline
53.250778 & $-$25.858597  &  53.284923 & $-$25.842899  \\
53.253821 & $-$25.865656  &  53.285166 & $-$25.909974  \\
53.257511 & $-$25.877557  &  53.285276 & $-$25.829389  \\
53.258258 & $-$25.874291  &  53.285467 & $-$25.843914  \\
53.259036 & $-$25.874262  &  53.286724 & $-$25.901295  \\
53.259263 & $-$25.872571  &  53.286868 & $-$25.839049  \\
53.259365 & $-$25.863121  &  53.287385 & $-$25.900560  \\
53.259620 & $-$25.871988  &  53.292858 & $-$25.831875  \\
53.260307 & $-$25.857930  &  53.293491 & $-$25.909665  \\
53.260432 & $-$25.869042  &  53.296294 & $-$25.910816  \\
53.260570 & $-$25.849192  &  53.296412 & $-$25.902661  \\
53.260858 & $-$25.857522  &  53.297229 & $-$25.903011  \\
53.261086 & $-$25.856151  &  53.297540 & $-$25.901051  \\
53.261435 & $-$25.872747  &  53.298007 & $-$25.904131  \\
53.261435 & $-$25.873184  &  53.298202 & $-$25.900421  \\
53.261694 & $-$25.874088  &  53.298940 & $-$25.907036  \\
53.262050 & $-$25.874438  &  53.304601 & $-$25.835236  \\
53.262308 & $-$25.877093  &  53.305184 & $-$25.834361  \\
53.262641 & $-$25.855889  &  53.306489 & $-$25.886072  \\
53.263115 & $-$25.883054  &  53.309578 & $-$25.833591  \\
53.263153 & $-$25.873505  &  53.312558 & $-$25.906791  \\
53.264545 & $-$25.878493  &  53.313039 & $-$25.828376  \\
53.264581 & $-$25.869394  &  53.318249 & $-$25.835481  \\
53.264824 & $-$25.889390  &  53.318716 & $-$25.834781  \\
53.265462 & $-$25.852157  &  53.321827 & $-$25.837476  \\
53.266691 & $-$25.888690  &  53.323609 & $-$25.910711  \\
53.266723 & $-$25.860586  &  53.323648 & $-$25.912286  \\
53.266846 & $-$25.891315  &  53.324621 & $-$25.915541  \\
53.267803 & $-$25.831625  &  53.328940 & $-$25.911550  \\
53.268862 & $-$25.862016  &  53.329290 & $-$25.912425  \\
53.269375 & $-$25.890336  &  53.329990 & $-$25.912845  \\
53.270156 & $-$25.881972  &  53.332752 & $-$25.906545  \\
53.270699 & $-$25.887222  &  53.333025 & $-$25.833765  \\
53.270846 & $-$25.842930  &  53.333219 & $-$25.907595  \\
53.271350 & $-$25.847200  &  53.333569 & $-$25.907910  \\
53.271391 & $-$25.839921  &  53.333958 & $-$25.907980  \\
53.271625 & $-$25.836596  &  53.336254 & $-$25.909064  \\
53.272206 & $-$25.846011  &  53.337306 & $-$25.913474  \\
53.272520 & $-$25.836246  &  53.337539 & $-$25.845139  \\
53.272839 & $-$25.886067  &  53.339485 & $-$25.915958  \\
53.274850 & $-$25.846606  &  53.343879 & $-$25.849057   \\
53.275052 & $-$25.899822  &  53.344965 & $-$25.840657   \\
53.275212 & $-$25.869192  &  53.347458 & $-$25.900767   \\
53.275380 & $-$25.851635  &  53.351033 & $-$25.845345   \\
53.275433 & $-$25.845592  &  53.351425 & $-$25.854795   \\
53.276572 & $-$25.889603  &  53.353602 & $-$25.892856   \\
53.277853 & $-$25.900418  &  53.358150 & $-$25.845763   \\
53.277966 & $-$25.827673  &  53.358930 & $-$25.851433   \\
53.279916 & $-$25.897724  &  53.359320 & $-$25.890019   \\
53.281116 & $-$25.853094  &  53.359712 & $-$25.861127   \\
53.281309 & $-$25.857702  &  53.362857 & $-$25.882318   \\
53.281936 & $-$25.909239  &  53.362947 & $-$25.876386   \\
53.282085 & $-$25.837718  &  53.363364 & $-$25.885398   \\
53.282477 & $-$25.853911  &  53.363565 & $-$25.866621   \\
53.282523 & $-$25.896709  &  53.363795 & $-$25.892852   \\
53.282872 & $-$25.899649  &  53.364033 & $-$25.870540   \\
53.283172 & $-$25.847903  &  53.365278 & $-$25.870295   \\
53.283223 & $-$25.899964  &  53.365864 & $-$25.875300   \\
53.283601 & $-$25.839644  &  53.368543 & $-$25.866549   \\
53.284342 & $-$25.831559  &  53.371654 & $-$25.865358   \\
\hline     
\end{tabular}
\end{table}

\end{document}